 \documentstyle[11pt, aas2pp4]{article}
\begin{document}
\newcommand{\cc}{\mbox{cm$^{-3}$}}
\newcommand{\tauv}{\mbox{$\tau_V$}}
\newcommand{\ra}{\mbox{$\rightarrow$}}
\newcommand{\nhtwo}{\mbox{n$_{H_{2}}$}}

\def\HI{H{\smc I}}
\def\HII{H{\smc II}}
%                                      molecules
\def\kms{km s$^{-1}$}               % M 17    
\def\m17{M~17}               % M 17    
\def\cepa{Cepheus~A}               %  Ceph A    
\def\Htwo{H$_2$}               % H2    
\def\HtwoO{H$_2$O}             % H2O 
\def\HtwoeiO{H$_2^{18}$O}             % H2(18)O 
\def\HtwoCO{H$_2$CO}           % H2CO 
\def\HtwoCS{H$_2$CS}           % H2CS 
\def\Hthreep{H$_3^+$}          % H3+
\def\HthreeOp{H$_3$O$^+$}          % H3+
\def\HCOp{HCO$^+$}             % HCO+
\def\HthCOp{H$^{13}$CO$^+$}    % H13CO+
\def\HtwCsiOp{H$^{12}$C$^{16}$O$^+$} % H12C16O+
\def\HCSp{HCS$^+$}             % HCS+
\def\HthCN{H$^{13}$CN}         % H13CN 
\def\HCfiN{HC$^{15}$N}         % HC15N 
\def\HtwCfoN{H$^{12}$C$^{14}$N}  % H12C14N 
\def\HNthC{HN$^{13}$C}         % HN13C
\def\HfoNtwC{H$^{14}$N$^{12}$C}  % H14N12C
\def\HCthreeN{HC$_3$N}         % HC3N 
\def\twCO{$^{12}$CO}           % 12CO
\def\thCO{$^{13}$CO}           % 13CO
\def\CseO{C$^{17}$O}           % C17O
\def\CeiO{C$^{18}$O}           % C18O
\def\twCsiO{$^{12}$C$^{16}$O}  % 12C16O
\def\thCsiO{$^{13}$C$^{16}$O}  % 13C16O
\def\twCeiO{$^{12}$C$^{18}$O}  % 12C18O
\def\thCeiO{$^{13}$C$^{18}$O}  % 13C18O
\def\CtfS{C$^{34}$S}           % C34S
\def\thCS{$^{13}$CS}           % 13CS 
\def\twCttS{$^{12}$C$^{32}$S}  % 12C32S
\def\tfSO{$^{34}$SO}           % 34SO
\def\ttSsiO{$^{32}$S$^{16}$O}  % 32S16O
\def\SOtwo{SO$_2$}             % SO2 
\def\tfSOtwo{$^{34}$SO$_2$}    % 34SO2
\def\SiO{SiO}             % SiO 
\def\Ntwo{N$_2$}               % N2
\def\Otwo{O$_2$}               % O2
\def\NtwoHp{N$_2$H$^+$}        % N2H+ 
\def\NHthree{NH$_{3}$}         % NH3
\def\CHthreeCCH{CH$_3$C$_{2}$H}     % CH3CCH
\def\CHthreeCN{CH$_3$CN}       % CH3CN
\def\CHthreeOH{CH$_3$OH}       % CH3OH
\def\CHfour{CH$_4$}       % CH4
\def\COtwo{CO$_2$}       % C02
\def\thCHthreeOH{$^{13}$CH$_3$OH}       % 13CH3OH
\def\twCHthsiOH{$^{12}$CH$_3$$^{16}$OH} % 12CH316OH
\def\CtwoH{C$_2$H}             % C2H
\def\CtwoS{C$_2$S}             % C2S
\def\CHp{CH$^{+}$}             % CH+
\def\Cp{C$^+$}             % C+
\def\CthreeHtwo{C$_3$H$_2$}    % C3H2
%                                      J transitions
\def\Jthoh{$J = 3/2 \to 1/2$}
\def\Johoh{$J = 1/2 \to 1/2$}
\def\Jtwel{$J = 12 \to 11$}
\def\Jelt{$J = 11 \to 10$}
\def\Jtn{$J = 10 \to 9$}
\def\Jne{$J = 9 \to 8$}
\def\Jes{$J = 8 \to 7$}
\def\Jss{$J = 7 \to 6$}
\def\Jsf{$J = 6 \to 5$}
\def\Jff{$J = 5 \to 4$}
\def\Jft{$J = 4 \to 3$}
\def\Jtt{$J = 3 \to 2$}
\def\Jto{$J = 2 \to 1$}
\def\Joz{$J = 1 \to 0$}
%                                      symbols
\def\WCO{W({\rm CO})}
\def\Wtw{W({\rm ^{12}CO})}
\def\Wth{W({\rm ^{13}CO})}
\def\dv{\Delta v}
\def\dvtw{\Delta v({\rm ^{12}CO})}
\def\dvth{\Delta v({\rm ^{13}CO})}
\def\NCO{N({\rm CO})}
\def\Nth{N({\rm ^{13}CO})}
\def\Ntw{N({\rm ^{12}CO})}
\def\NtwCsiO{N({\rm ^{12}C^{16}O})}
\def\NthCO{N({\rm ^{13}CO})}
\def\NthCsiO{N({\rm ^{13}C^{16}O})}
\def\NtwCeiO{N({\rm ^{12}C^{18}O})}
\def\intCO{\int T_R({\rm CO})dv}
\def\inttwCsiO{\int T_R({\rm ^{12}C^{16}O})dv}
\def\intthCsiO{\int T_R({\rm ^{13}C^{16}O})dv}
\def\inttwCeiO{\int T_R({\rm ^{12}C^{18}O})dv}
\def\NHtwo{N({\rm H_2})}
\def\Wtw{W_{12}}
\def\Wth{W_{13}}
\def\kappanu{\kappa_{\nu}}
\def\phinu{\varphi_{\nu}}
\def\taunu{\tau_{\nu}}
\def\dv{\Delta v}
\def\dvFWHM{\Delta v_{FWHM}}
\def\vLSR{v_{LSR}}
\def\Rsol{R_\odot}
\def\Msol{M_\odot}
\def\MMsol{\ts 10^6\ts M_\odot}
\def\MCO{M_{\rm CO}} 
\def\Mvir{M_{\rm vir}}
\def\TAstar{T^*_A}
\def\TAstartwCO{T^*_A(^{12}{\rm CO})}
\def\TAstarthCO{T^*_A(^{13}{\rm CO})}
\def\TAstarCeiO{T^*_A({\rm C}^{18}{\rm O})}
\def\TRstar{T^*_R}
\def\TexCO{T_{ex}({\rm CO})}
\def\Trms{T_{rms}}
%                                      			units
\def\d{^\circ}
\def\h{^{\rm h}}
\def\mi{^{\rm m}}
\def\s{^{\rm s}}
\def\mum{\ts \mu{\rm m}}
\def\mm{\ts {\rm mm}}
\def\cm{\ts {\rm cm}}
\def\percm{\ts {\rm cm}^{-1}}
\def\m{\ts {\rm m}}
\def\kms{\ts {\rm km\ts s^{-1}}}
\def\K{\ts {\rm K}}
\def\Kkms{\ts {\rm K\ts km\ts s^{-1}}}
\def\kHz{\ts {\rm kHz}}
\def\MHz{\ts {\rm MHz}}
\def\GHz{\ts {\rm GHz}}
\def\pc{\ts {\rm pc}}
\def\kpc{\ts {\rm kpc}}
\def\Mpc{\ts {\rm Mpc}}
\def\cmsq{\ts {\rm cm^2}}
\def\pcsq{\ts {\rm pc^2}}
\def\dsq{\ts {\rm deg^2}}
\def\debye{\ts10^{-18}\ts {\rm esu}\ts {\rm cm}}

%                                      			journals
%                                      			mathe
\let\ap=\approx
\let\ts=\thinspace

\title{THE POST-SHOCK CHEMICAL LIFETIMES OF OUTFLOW TRACERS AND A POSSIBLE NEW MECHANISM TO 
PRODUCE WATER ICE MANTLES}

\author{Edwin A. Bergin$^{1}$, Gary J. Melnick$^{1}$, and
David A. Neufeld$^{2}$}

\noindent$^1$ {Harvard-Smithsonian Center for Astrophysics, MS-66, 
60 Garden St., Cambridge, MA 02138; 
ebergin, gmelnick@cfa.harvard.edu}\newline
\noindent$^2$ {Department of Physics and Astronomy, The Johns Hopkins University,
3400 North Charles Street, Baltimore, MD 21218; neufeld@pha.jhu.edu}

\begin{abstract}
We have used a coupled time-dependent chemical and dynamical model to investigate
the lifetime of the chemical legacy left in the wake of C-type shocks.  
 We concentrate this study on the chemistry of \HtwoO\ and \Otwo , two molecules
 which are predicted to have abundances that are significantly affected in 
 shock-heated gas.
Two models
are presented: (1) a three-stage model of pre-shock, shocked, and post-shock gas; and
(2) a Monte-Carlo cloud simulation where we explore the effects of stochastic shock
activity on molecular gas over a cloud lifetime.   For both models we separately examine
the pure gas-phase chemistry as well as the chemistry including the interactions of molecules
with grain surfaces.  In agreement with previous studies, we find that shock velocities
in excess of 10 km s$^{-1}$ are required to convert all of the oxygen not locked
in CO into
\HtwoO\ before the gas has an opportunity to cool.   For pure gas-phase models the
lifetime of the high water abundances, or ``\HtwoO\ legacy'', in the post-shock gas is
$\sim 4 - 7 \times 10^{5}$ years, independent of the gas density.  A density dependence
for the lifetime of \HtwoO\ is found in gas-grain models as the water molecules deplete
onto grains at the depletion timescale.  

Through the Monte Carlo cloud simulation we demonstrate that the time-average 
abundance of \HtwoO\ -- the weighted average of the amount of time gas spends in pre-shock,
shock, and post-shock stages -- is a sensitive function of the frequency of shocks.
Thus we predict that the abundance of \HtwoO , and to a lesser extent \Otwo , can
be used to trace the history of shock activity in molecular gas.  We use previous large-scale
surveys of molecular outflows to constrain the frequency of 10 km s$^{-1}$ shocks in
regions with varying star-formation properties and
discuss the observations required to test these results.
We discuss the post-shock lifetimes for other possible outflow tracers
(e.g. SiO, \CHthreeOH) and show that the differences between the lifetimes
for various tracers can produce potentially observable chemical variations between 
younger and older outflows. 
For gas-grain models we find that the abundance of water-ice on grain surfaces
can be quite large and is comparable to that observed in molecular clouds.  This
offers a possible alternative method  to create water mantles without resorting to
grain surface chemistry: gas heating and chemical modification due to a C-type
shock and subsequent depletion of the gas-phase species onto grain mantles. 

\end{abstract}


\lefthead{}
\righthead{}

\pagebreak

\section{Introduction}

The importance of molecular gas as the material from which stellar and planetary
systems are made has led numerous investigators to examine the chemical processes
that combine atoms into molecules.  Over the past two decades these studies
have gradually increased in both complexity and fidelity and can roughly
be divided into three generations.  The first generation chemical model
solved the chemical rate equations at equilibrium, or steady-state, and
demonstrated the importance of ion-molecule reactions in driving 
the gas-phase chemistry (\cite{HK73}).  The second generation of models
utilized the later availability of increased computing power to study
the time evolution of chemical abundances 
(\cite{PH80}; \cite{GLF82}).
These models, labeled as ``pseudo-time dependent'' -- 
``pseudo'' because the chemistry evolves with fixed physical conditions --
used observed physical properties of dense molecular cores 
($T_k \sim 10 - 30$ K, \nhtwo\ $= 10^{4-6}$ \cc ) to show that
chemical equilibrium was reached in $\sim 10^{6-8}$ years. 
Pseudo-time dependent models represent the most prevalent chemical model
and have been quite successful in describing
the chemistry of quiescent regions in both dark and giant molecular cloud
cores (\cite{Lee_etal96}; \cite{BGSL97}). 

Second generation chemical models operate 
under one simplifying assumption: that the molecular
gas undergoes no dynamical evolution as it chemically evolves.
However, molecular clouds are certainly dynamically evolving objects.  
The widespread occurrence of superthermal widths in molecular
lines suggests that, overall, the evolution of molecular clouds is not
simple quiescent evolution at a single gas
temperature.  Moreover, the formation of a low- or
high-mass star from a molecular condensation involves a collapse
that increases the density by many orders of magnitude.  It has also
been recognized that the birth of a protostar is associated with a
period of intense mass loss, which manifests itself in energetic winds,
bipolar jets, and large-scale outflows (c.f. \cite{Lada85}; 
\cite{Bachiller96}).  The impact of energetic flows on surrounding
quiescent gas can compress and heat
the gas, in some cases providing enough energy to overcome endothermic barriers
of chemical reactions, sputter or destroy grains,
or even dissociate molecules.  The result can be a considerably
different chemical composition in the shocked
gas than observed in quiescent material.  
Since molecules are the primary coolants of the gas, any chemical
change induced by collapse or shocks can also alter the ensuing physical
evolution of the cloud.
To address such concerns, a third generation of chemical models has
been constructed to build on the successes of previous generations
by combining
chemistry with dynamics.  These models have been principally directed towards 
investigating the chemistry of core
and star formation (c.f. \cite{PTVH87}; \cite{RHMW92}; \cite{BL97}), 
the physical and chemical structure of shocked gas (c.f. \cite{DM93}), or even
complex scenarios in which gas is cycled between low and high density through 
recurrent episodes of low-mass star formation (\cite{CDHW88a}).

One goal of coupled astrochemical
models is to search for and isolate specific molecular species that serve as 
signposts of particular dynamical events, such as shocks. 
Coupled models of shocked molecular gas have isolated one molecule in
particular, \HtwoO , which is predicted to form in large abundance in
shock-heated gas (\cite{DRD83}; \cite{KN96a},b);
if temperatures in the shocked gas exceed 400~K, 
the endothermic barriers of a few key chemical reactions are exceeded and 
all of the available oxygen not locked up in CO will be driven into \HtwoO . 
This led to the assertion that water emission is a ubiquitous tracer of
shock-heated gas (\cite{NM87}).  Unfortunately, due to constraints imposed
by the earth's atmosphere, the detection of \HtwoO\ in
interstellar clouds from ground-based observatories is a challenging task. 
Nevertheless, several studies have observed, and detected, transitions of
isotopic water (\HtwoeiO), and even \HtwoO\
(c.f. \cite{Jacq_etal88}; \cite{KL91}; \cite{Zmuidzinas_etal94}). 
Quite recently, absorption from warm ($T_{gas} > 200$ K) water vapor
has been unambiguously detected by the Infrared Space Observatory (ISO)
towards hot stars (\cite{Helmich_etal96};
\cite{vDH96}), and in emission in HH54 (\cite{Liseau_etal96}).  The
inferred water abundance in these sources is $\sim 1 - 6 \times 10^{-5}$, 
much larger than predicted by chemical models run for cold quiescent
conditions ($x$(\HtwoO ) $\sim$10$^{-7}$),
and is consistent with water production in warm gas either though 
high-temperature chemistry -- as would apply in shock-heated gas or
in the near vicinity of embedded sources -- or via evaporation of water-ice mantles.

An important question yet to be addressed by coupled dynamical and
chemical models of shocked gas in molecular clouds is how long the
high abundances of water and other molecules persist following
the passage of a shock.   As we will demonstrate,
the time needed for shock-heated gas to 
return to its pre-shock temperature is several orders of magnitude
less than the time required for the gas to return to its pre-shock
chemical composition.  Thus, the enhanced abundances of
water and other species could potentially exist long after the dynamical
effects of a shock passage are dissipated. Therefore, 
the chemistry inside a cloud is reflective of, and can be used to probe, 
the physical shock history of the molecular gas.

Motivated by these questions, we present the results of a coupled
dynamical and chemical study of molecular gas that is subjected
to shocks with velocities greater than 10 km s$^{-1}$.  
In particular, we follow the time dependence of the chemistry in
a shock-heated gas layer as it cools and the
quiescent time-dependent chemistry is ultimately
re-established.  
We present models examining this 
evolution using pure gas-phase chemistry as well as chemistry which includes
the interaction of molecules with grain surfaces.    

In Section 2 we discuss the combined chemical and dynamical model.
In Section 3 we use published surveys of molecular outflows to investigate the
average rate at which shocks with a minimally sufficient velocity to affect
the water chemistry -- 10 km s$^{-1}$ -- pass a given region of a cloud.
Section 4 presents the results from our combined models along with
one example of quiescent chemical evolution.  Two models are presented:
(1) a three-stage model of pre-shock, shocked, and post-shock gas;
and, (2) a Monte-Carlo cloud simulation
in which the results from Section 3 are used to examine the effects of
stochastic shock activity on molecular gas over a cloud lifetime.
Section 5 discusses the importance of these results on chemical models
of molecular gas and also reviews the observations required to test
these models.   In Section 6 we summarize our results.

\section{Dynamical and Chemical Model}

\subsection{Shocks and Chemistry}
\begin{figure*}
\figurenum{1}
\plotfiddle{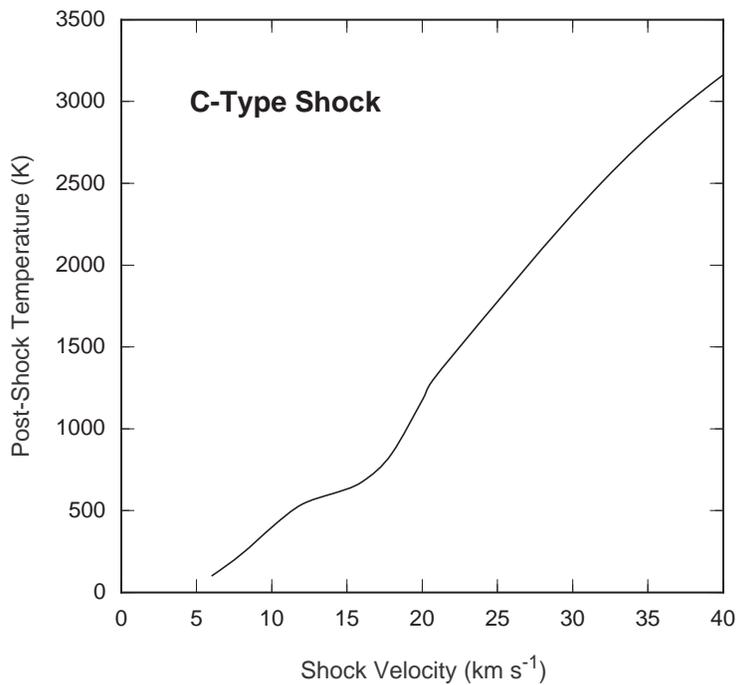}{2.5in}{0}{60}{60}{-210}{-30}
\caption{Maximum attained gas temperature as a function of shock velocity for 
density of $n_{H_2} = 10^5$ cm$^{-3}$ (taken from Kaufman \& Neufeld 1996b).  
The slight drop in temperature at $\sim$15 km s$^{-1}$ is due to the 
onset of H$_2$O formation and cooling.
}
\label{shocktemp}
\end{figure*}

Detailed models of magnetohydrodynamic (MHD) shocks in dense molecular clouds 
have been created by a number of authors.  For a comprehensive discussion of
various models the reader is directed to the review in Draine \& McKee (1993). 
Briefly, MHD shocks in interstellar gas are divided into two categories:
(1) J-type shocks where flow variables are discontinuous across the shock;
and, (2) C-type shocks where the flow variables are continuous across the
shock front.  
For J-shocks the energy dissipation occurs in a thin layer where the 
temperature increase is high enough to destroy molecules.   
In C-type shocks propagating through a magnetic medium with low ionization
(the case for dense molecular gas) the gas is frictionally
heated in a thicker layer through the process of ambipolar diffusion
wherein ions tied to the field drift through the neutral fluid.   
The temperatures produced by this process are as high as several
thousand Kelvins, but the gas remains in molecular form (\cite{Mullan71}; 
\cite{Draine80}).  
In addition, the gas compression in the wake of C-shocks in the region of the
temperature peak is not exceedingly
large (c.f. \cite{DR82}); to first order, the neutral density in the gas which will
be most affected by the temperature increase is approximately 
constant.  The magnitude of the heating depends on
the pre-shock density, the strength of the magnetic field, and the velocity
of the shock.  For the purposes of this paper we are mostly concerned with
the non-dissociative C-type shocks that drive observed molecular outflows.
We account for the possibility of dissociative shocks 
in the more comprehensive cloud model presented in Section 4.4.

\begin{figure*}
\figurenum{2}
\plotfiddle{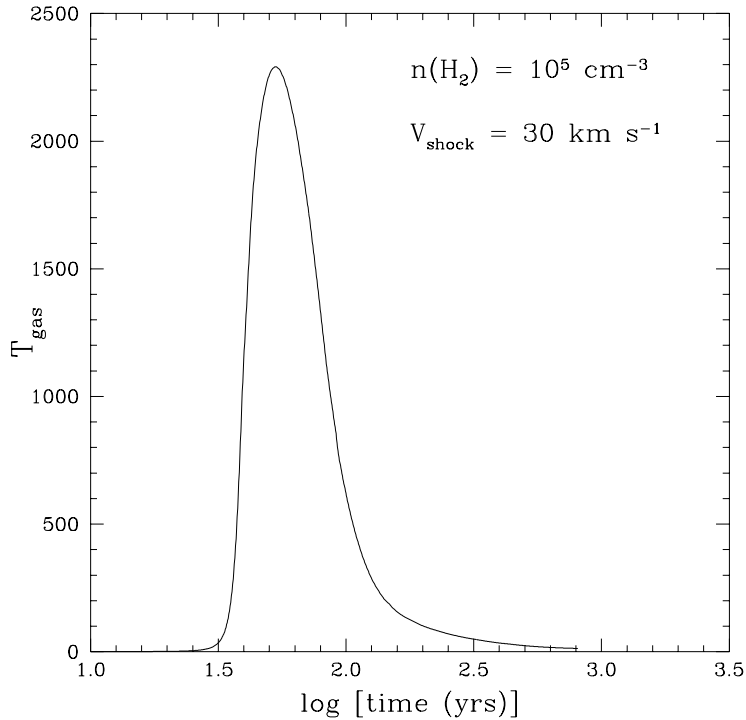}{2.5in}{0}{50}{50}{-150}{-80}
\caption{Shock temperature as a function of time for $v_s = 30$ km s$^{-1}$ and
$n_{H_2} = 10^5$ cm$^{-3}$ (provided by M. Kaufman).
}
\label{coolingtime}
\end{figure*}
In the models shown here we are solely concerned with the chemistry in
the post-shock gas, which is largely driven by the gas temperature, molecular
hydrogen density, and the visual extinction.  To simplify the computational
task, we adopt many of these conditions from detailed MHD shock models (e.g.,
Kaufman and Neufeld 1996a,b: hereafter KN96a and KN96b), including the variations in these
parameters with both time and shock velocity.  For C-shocks, the post-shock
\Htwo\ density in the region of the temperature maximum
can be held constant.  In addition, most shock models assume
that the post-shock gas is shielded from dissociating radiation, an assumption
we adopt as well.    Thus, the most important remaining parameters to be
extracted from the MHD shock models are the velocity-dependent temperature
profiles and their time evolution.

KN96b present a self-consistent calculation of the heating,
cooling, and chemistry
of a MHD shock wave propagating through dense (\nhtwo\ = 10$^{4-6.5}$ \cc )
molecular gas with low ($< 10^{-7}$) ionization.  The magnetic
field in their model is assumed to be perpendicular to the direction
of shock propagation with a value of $B = (n_H/cm^{-3})^{1/2}$ $\mu$G.
The assumed values for the ionization fraction and magnetic field 
strength are consistent with those estimated
in molecular clouds (e.g. Gu\'elin et al. 1982; Troland \& Heiles 1986).
In Figure 1 we show the maximum shock temperature
as a function of the shock velocity (adapted from KN96b).   
This dependence is shown for a pre-shock molecular hydrogen density of
$10^5$ \cc , but the maximum post-shock temperature is only
weakly dependent on the density (KN96b).  Clearly the gas temperature is a
strong function of the shock velocity, and is essentially a reflection of
the strength of the frictional heating created by ambipolar diffusion.

Another important parameter required to examine the chemistry of post-shock gas is
the amount of time the temperature remains at elevated levels.   In Figure 2
we present the temperature as a function of time during the passage of 
a 30 km s$^{-1}$ shock in a medium with a pre-shock density
n$_{\rm H_2} = 10^{5}$ \cc .
These data have been kindly provided by M. Kaufman using the code described in
KN96a,b.  The effect of the shock
is clearly evident as the gas temperature rises to greater than 2000 K in 
less than 100 years.   After the passage of the shock, the gas rapidly cools
within a few hundred years.
The principal coolants of the gas are \Htwo , \HtwoO , and CO.  
Thus, the passage of a 30 km s$^{-1}$ shock could heat the gas from
T$_{gas}$ = 10--30 K to $\sim 2000$ K, and this gas will return to
30 K within a few hundred years.

\subsection{Chemical Model}

In these calculations we have used the chemical model initially described in
Bergin, Langer, \& Goldsmith (1995) and updated in
Bergin \& Langer (1997).  These papers examined the gas-phase chemical evolution
including grain surface molecular depletion and desorption.  For this calculation
we have used both the gas-phase UMIST RATE95 chemical network (\cite{MFW97})
and the gas-grain adaption of this network described in Bergin \& Langer (1997).
Thus, we separately examine the pure gas-phase chemistry and the evolution
including grain processes.
The gas-grain models include depletion onto grain surfaces and desorption
via three desorption mechanisms: cosmic-ray-induced heating, photodesorption, and
simple thermal evaporation (see \cite{BLG95}).
For these models we use the binding energies of molecules on
an SiO$_2$ grain surface.

For a given calculation the adjustable variables are the density of molecular
hydrogen, \nhtwo , the factor by which the external UV field is enhanced above
the average galactic interstellar radiation field, $G_{\rm o}$, the gas temperature,  $T_{gas}$,
and the visual extinction to the center of the cloud, $A_V$.  For gas-grain
models, the dust temperature ($T_{dust}$) is also a variable.  The initial chemical
conditions are taken from Millar et al. (1991), and assume that the
carbon, silicon, iron, sulfur, sodium, potassium, and magnesium are
in ionized form, while the oxygen
and nitrogen atoms are neutral.  All species are assumed to be 
depleted relative to their solar abundances,
while the heavy atoms (e.g., Fe, Mg) are assumed to be
more severely depleted than carbon, oxygen, and nitrogen.  

For chemical models of interstellar shocks, it is particularly important
that the chemical network contains rate coefficients appropriate for
high-temperature chemistry.
To minimize such concerns, we only discuss in detail results for a few 
simple carbon- and oxygen-bearing species (C, CO, \HCOp , O, \Otwo , OH, and \HtwoO )
whose chemistry is well characterized at both high and low temperatures.
As mentioned earlier, when temperatures exceed $\sim$400 K, water is rapidly
formed in the gas phase through the following series of neutral-neutral reactions

\begin{equation}
\eqnum{R1}
\rm{O  +   H_2   \rightarrow   OH   +   H}
\end{equation}

\begin{equation}
\eqnum{R2}
\rm{OH + H_2 \rightarrow H_2O + H}
\end{equation}

\noindent (\cite{EdJ73}; \cite{EW78}).   
Both of these reactions possess significant activation
barriers ($E_a(R1) = 3160$ K; $E_a(R2) = 1660$ K). 
For very high temperatures,
molecular oxygen will be destroyed by reaction R3,

\begin{equation}
\eqnum{R3}
\rm{O_2 + H_2 \rightarrow OH + OH}
\end{equation} 

\noindent 
($E_a(R2) = 28190$ K),
and eventually processed into \HtwoO\ through 
reaction R2.
We examine in detail the temperature dependence and efficiency of reactions R1--3 
in the following section.

The chemical network includes many other species beyond the simple species
we emphasize in this paper and, unless otherwise noted, these species have been 
included in the model
runs.  Because we do not have the same confidence in the high-temperature
chemical reactions that form these species, we do not provide their predicted abundances.  
The UMIST database 
contains rate equations for gas temperatures between 10 -- 50 K
(Millar, Farquhar, \& Willacy  1997), a more consistent
calculation for other potentially important shock tracers, such as SiO or SO,
requires the incorporation of other reactions such as
those estimated for sulphur-bearing species in Pineau des For\^ets et al. (1993).
However, in Section 5.2 we do discuss the relevance these results might have on 
the abundances and post-shock lifetimes of some of these ``shock tracers.''

Careful attention has also been paid to the accuracy of the low-temperature reaction
rates, although we have sought to minimize this concern through the use of
the UMIST database.
Of particular importance for the abundance of water at low temperatures
are two reactions that have
been the subject of considerable debate.  These are the 
dissociative recombination rate of \Hthreep , and the branching ratios of the
\HthreeOp\ recombination reaction.  For \Hthreep, the UMIST RATE95 database has adopted
a rate of $\alpha(H_3^+) = 3 \times 10^{-8}T^{-0.5}$, from Smith and Spanel (1993).
This rate has been slightly revised by the recent measurements of Sundstr\"{o}m et
al. 1994.  The adoption of the more recent value will not alter our results. 
Recently, there have been two studies of the \HthreeOp\ dissociative recombination
branching ratios (\cite{Williams_etal96}; \cite{Vejby-Christensen_etal97}).  
The branching paths are:

\begin{mathletters}
\eqnum{R4}
\begin{eqnarray}
\rm{H_3O^+ + e^-}  &\rightarrow & \rm{H + H_2O}\;\;\;\;\;\;\;\;\;\;\: f_1\\
& \rightarrow & \rm{OH + H_2}\;\;\;\;\;\;\;\;\;\;\:  f_2 \nonumber\\
&\rightarrow & \rm{OH + 2H}\;\;\;\;\;\;\;\;\;\;\; f_3 \nonumber\\
&\rightarrow & \rm{O + H + H_2}\;\;\;\;\;\;\;f_4\nonumber
\end{eqnarray} 
\end{mathletters}

\noindent where $f_{1-4}$ are the branching ratios.   Vejby-Christensen et al. (1997) 
using an ion storage ring measured $f_1 = 0.33$, $f_2 = 0.18$, $f_3 = 0.48$, and
$f_4 = 0.01$, while Williams et al. (1996) used an ion flow tube to measure
$f_1 = 0.05$, $f_2 = 0.36$, $f_3 = 0.29$, $f_4 = 0.3$.  The overall OH production
predicted by these two studies are in reasonable agreement,
which is important since OH reacts with O to form \Otwo .  
However, the measurements differ significantly in their
production of water and atomic oxygen.   For the following calculations
we have used the branching ratios estimated by
Vejby-Christensen et al. (1997).
In order to examine the sensitivity of these results to this choice,
we present one case using the other branching ratios.

\section{Shock Timescales in Molecular Clouds}

Before we examine the timescale for the re-assertion of quiescent chemistry after the
passage of a shock, it
is useful to examine how often molecular material in molecular cores is subjected
to 10 km s$^{-1}$ shocks.  The choice of this particular shock velocity
is discussed in Section 4.2.  To estimate the rate at which a parcel of
molecular gas might be subjected to a 10 km s$^{-1}$ shock, we have used a
method outlined in Margulis, Lada, and Snell (1988).   Margulis and Lada (1986)
surveyed a $\sim 0.5^{\circ} \times 2^{\circ}$
region of the Mon OB1 molecular cloud in the emission of $^{12}$CO and identified
nine separate molecular outflows.  To examine the effect of the molecular outflows
on the dynamics of the entire molecular core, Margulis et al. (1988) estimate the
total momentum added to the cloud by the outflows.  The momentum of an outflow is
estimated by either: (a) integrating directly over the mass $\times$ velocity
in each velocity channel in the line wings, ignoring the possible contribution
of outflow material hidden by the line core; and/or, (b) 
assuming that the entire outflow is at a constant velocity, which is equated to the
maximum observed flow velocity (c.f. Lada 1985).  
Using method (a) provides a lower limit to
the momentum, because it ignores the inclination of the outflow and, because of
the difficulty
of differentiating the line wings from the core, also ignores 
the mass in the line core.  In contrast, method (b) is a strict upper limit, because
it assumes that all of the mass in the flow is at the highest observed velocity.

Using method (a) and the momenta listed in Table 2 of Margulis et al. (1988),
the total momentum generated by the observed outflows in Mon OB1 is 
163 M$_{\odot}$ km s$^{-1}$.  Momentum conservation requires that 16 M$_{\odot}$
of molecular material will have been swept up when the outflows have reached
a velocity of 10 km s$^{-1}$.  Since the 
total mass in the portion of Mon OB1 surveyed is 3 $\times 10^4$ M$_{\odot}$
(Margulis and Lada 1986), only 0.05 percent of the total mass will be affected
by the energetic
flows.  Thus, in the momentum conserving case, it requires $\sim$1900 generations of 
similar star-formation activity for the entire surveyed area to have experienced a
shock of $\sim$10 km s$^{-1}$.  
Assuming that the average duration of the outflow phase is $\sim 2 \times
10^4$ year (Staude and Els\"{a}sser 1993), then an upper limit to the shock timescale,
$\tau_{s}$, for gas in Mon OB1 to be hit by 10 km s$^{-1}$ shocks
is $< 4 \times 10^7$ years.  Because
the outflow momentum estimates are strict lower limits (see discussion of 
method (a) above), and due to the assumption
that subsequent generations of star-formation activity occurs continuously, 
this timescale is an upper limit.
Similarly, we can use the momenta calculated via
method (b) to set a lower limit for this timescale.  Using method (b), Margulis et al.
(1988) calculate the total momentum to be 883  M$_{\odot}$ km s$^{-1}$, which is
3 percent of the total mass of Mon OB1.  Therefore, $\sim$340 generations of
similar star-formation activity are required to process the entire core by 10 km s$^{-1}$
shocks, and the lower limit to the timescale is 
$\tau_{s}$(Mon OB1) $> 7 \times 10^6$ year.   
We have performed a similar analysis 
using the observations of L1641 in Morgan et al. (1991) and find similar results,
i.e., $\tau_s$(L1641)$ \sim 0.6 - 2.7  \times 10^7$ years. 

\begin{figure*}
\figurenum{3}
\plotfiddle{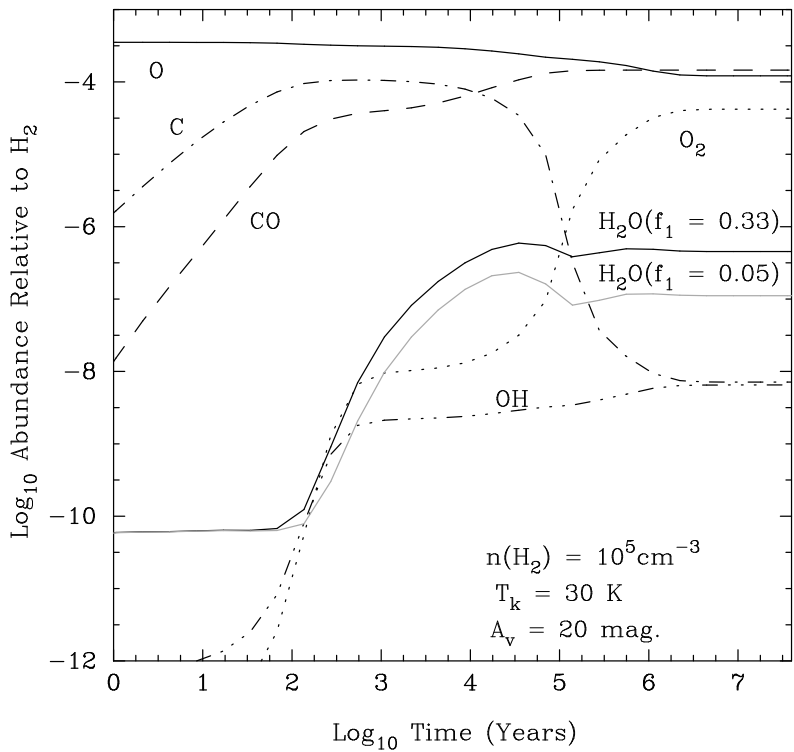}{2.5in}{0}{65}{65}{-190}{-185}
\caption{
Time evolution of chemical abundances relative to H$_2$ for quiescent physical conditions.  The
evolution of the H$_2$O abundance is shown for two values of the H$_3$O$^{+}$ 
branching ratio (discussed in Section 3).
}
\label{Qchem}
\end{figure*}
It's important to note that this ``cloud-average'' approach is useful for investigating
the potential of shocks to alter the global composition of a cloud.  Clearly  though,
this is an over-simplification.  In reality, an observer will see portions of a cloud that
are experiencing active star formation, for which the local value of $\tau_s$ may
be $\ll 10^6$ years, while more quiescent regions of the same cloud may have local
values of $\tau_s$ closer to a few times $10^7$ yrs, the upper limit quoted above.
For instance, NGC 1333 is an example of a molecular core where a cluster
of outflows has been detected within
a $15' \times 15'$ (1.5 $\times$ 1.5 pc) area (see Liseau et al. 1998, 
Warin et al. 1996 and references therein).   Thus, {\em locally} the shock timescale
in NGC 1333 may be below the global value of $\sim 10^{7}$ years.
Another example is the Orion Nebular Cluster
where several thousand stars are estimated to have been formed within the past
1-2 million years (Hillenbrand 1997).  If each new star has an an associated outflow
phase, it is certainly possible that the shock timescale {\em in the dense core associated
with this cluster (OMC-1)} can be reduced to values 
well below the global value.    
As will be shown, a range in the values of $\tau_s$ between $< 10^6$ and $> 10^7$ years
spans a large range in chemical outcomes and may ultimately prove to be the cause of
strong compositional gradients within clouds.

\section{Results}

In this section we present the results of the coupled dynamical and chemical
model of molecular gas that is subject to shocks with a range of
velocities.  For quiescent conditions, we assume, unless otherwise noted, that
\nhtwo\ $= 10^5$ \cc , $T_{gas} = T_{dust} = 30$ K, $G_{\rm o}$ = 1,
and $A_V = 20$ mag.
These conditions are similar to those observed in Giant Molecular Cloud (GMC)
Cores, the sites of massive star formation (\cite{Goldsmith87}).  

\subsection{Quiescent Chemistry}

In Figure 3 we present the time dependence of the gas-phase
chemistry for the quiescent conditions 
appropriate in GMC cores.  In this figure we also show the water abundance calculated
using two different values for the \HthreeOp\ dissociative recombination
branching ratios (see discussion in Section 2.2).  
The observed time evolution of chemical abundances is quite similar to 
other current gas-phase chemical models (\cite{Lee_etal96}; \cite{MFW97}).    
Examination of the evolution shows that the carbon abundance peaks at early
times ($t \sim 10^4$ years for \nhtwo\ = 10$^{5}$ \cc ), while for later times
all carbon is locked into CO.  Equilibrium is reached after $\sim 3 \times 10^6$ years, 
when the dominant oxygen reservoirs are O, CO, and \Otwo .
In addition, due to cosmic-ray-induced photodissociation of \Otwo\
(\cite{SD95}),
most oxygen remains
in atomic form, although the total amount of oxygen in \Otwo\ (2 $\times x(\rm{O_2})$)
is similar to the abundance of atomic oxygen.  
The steady-state water abundance is different for the two measured branching ratios;
for $f_1 = 0.33$ the abundance is $x(\rm{H_2O}) \sim 4.5 \times 10^{-7}$, while for
$f_1 = 0.05$, $x(\rm{H_2O}) \sim 1.0 \times 10^{-7}$.  For all other calculations we
use $f_1 = 0.33$.

\subsection{High-Temperature Chemistry}

%As discussed in Section 3, the abundance of \HtwoO\ and \Otwo\ 
%will be altered at high temperatures through 
%the influence of the neutral-neutral reactions, R1--3, discussed earlier.
In Figure 4 we present the dependence of the water abundance on both time
and temperature, while Figure 5 presents a similar plot for \Otwo .  
The models are for pure gas-phase chemistry
where we allow the chemical evolution to proceed for 10$^5$ years
with $T_{gas} = 30$ K, at which
point the chemistry is re-initiated and the temperature is increased to 
the values indicated in Figures 4 and 5.   This method allows for appreciable
abundances of molecular species to build up in the gas phase at low temperature
and makes it easier to gauge the
effects of high temperature chemistry.  Thus, $t = 0$ in these figures refers to 
time after 10$^{5}$ years of low-temperature or quiescent chemical evolution when
the abundances of water and molecular oxygen are
$x(\rm{H_2O}) \sim 4 \times 10^{-7}$ and $x(\rm{O_2}) \sim 3 \times 10^{-7}$.
The results presented in Figures 4 and 5 do not depend in detail
on the time the chemistry is allowed to evolve under quiescent conditions. 

In Figure 4, for $T_{gas} > 500$ K (corresponding to shock velocities, $v_s > 10$ 
km s$^{-1}$), water accounts for all of the oxygen not bound as CO
within $\sim$100 years. 
For temperatures below $\sim$400 K, the 
efficiency of reactions R1 and R2 begins to rapidly decline
until $T_{gas} \sim 200$ K
where the high-temperature reactions are unimportant compared to the
low-temperature formation pathways.    
In contrast, the abundance of \Otwo\ shown in Figure 5 exhibits the
opposite behavior.  For $T_{gas} = 400$ K the \Otwo\ abundance increases
above the initial value through normal time-dependent chemistry.
However, at $t \sim 2 \times 10^5$ years, the abundance of
\Otwo\ exhibits a rapid decrease
and reaction R3 effectively processes \Otwo\ into \HtwoO .  For even higher
temperatures the destruction of \Otwo\ becomes more efficient, although it
requires gas temperatures greater than 2000 K, corresponding to $v_s \geq 26$ km s$^{-1}$,
to destroy all \Otwo\ within 100 years.

As demonstrated in Section 2, the post-shock gas will not
remain at elevated temperatures for the entire time shown in Figures 4 and 5.
Using the cooling information provided in Figure 2, we see that 
the time for the post-shock gas to cool to below 200 K, what we term
the cooling timescale, is $\sim 100$ years.
Combining the information shown in Figures~2 and
4 suggests that {\em shocks with velocities $>$ 10 km s$^{-1}$ will convert all of the
oxygen to water within a cooling timescale}.  The maximum temperatures
reached behind
shocks with velocities less than 10 km s$^{-1}$ will not be high enough to
overcome the activation barriers and the abundance of water should remain
relatively unaltered from its quiescent values.
This result is similar to that found by KN96a,b who found that the maximum
shock velocity for the conversion of O to H$_2$O is 10 -- 15 km s$^{-1}$.
Similarly, shock velocities $>$ 26 km s$^{-1}$ are required to destroy all 
molecular oxygen within the cooling timescale.
While the exact cooling timescale will vary with shock velocity, the cooling will
be most efficient when the oxygen is mainly in the form of water -- thus, the cooling
timescale should not vary appreciably as long as $T_{gas} > 400$ K.

\begin{figure*}
\figurenum{4}
\plotfiddle{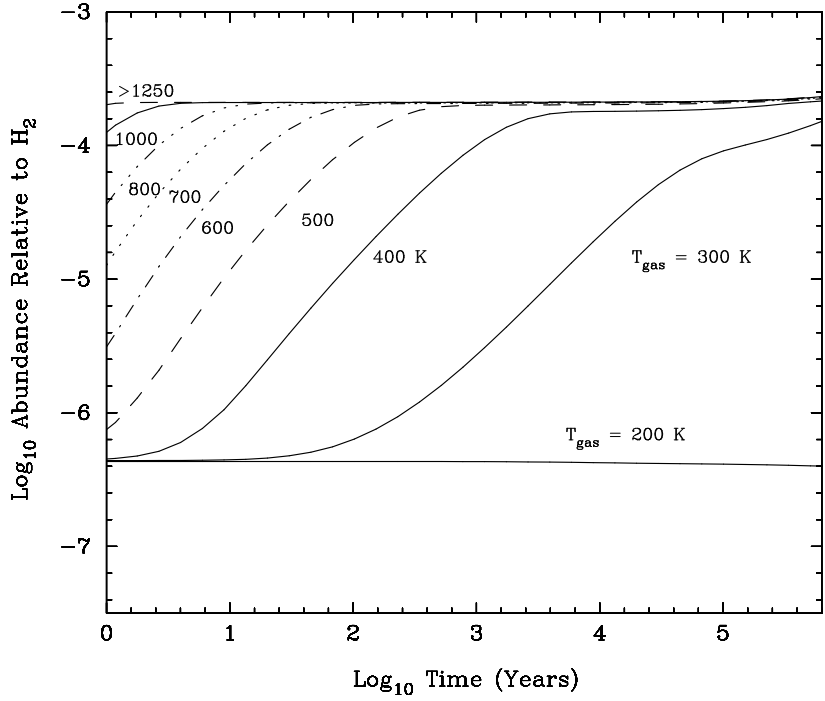}{2.5in}{0}{60}{60}{-200}{-170}
\caption{
Abundance of H$_2$O as a function of time and gas temperature.  The times listed on
the abscissa are times after $\sim 10^{5}$ years of evolution with $T_{gas} = 30$ K
(see Section 4.1).  
}
\label{h2otemp}
\end{figure*}

\begin{figure*}
\figurenum{5}
\plotfiddle{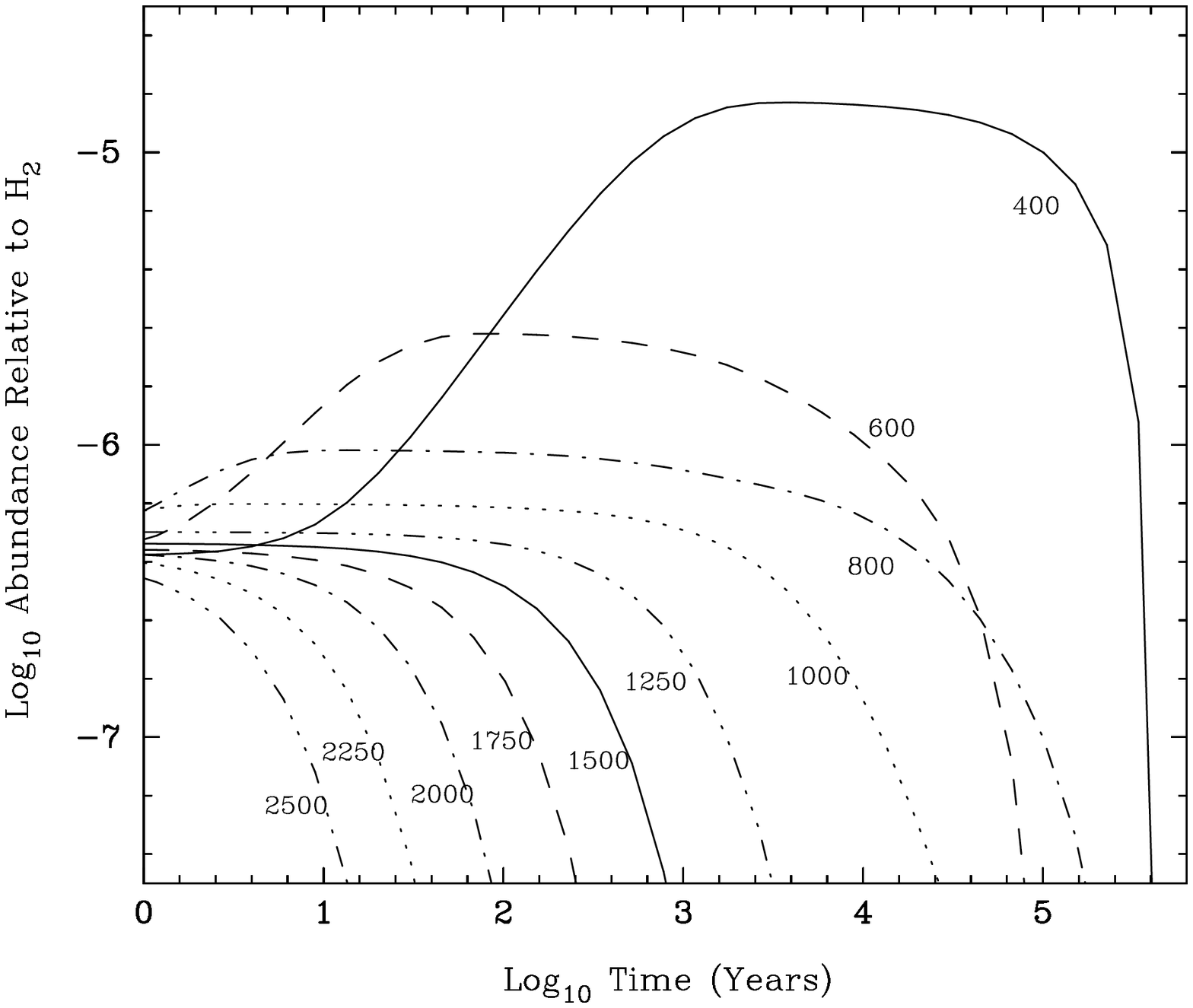}{2.5in}{0}{60}{60}{-200}{-170}
\caption{
Abundance of O$_2$ as a function of time and gas temperature.  The times listed on
the abscissa are times after $\sim 10^{5}$ years of evolution with $T_{gas} = 30$ K
(see Section 4.1).  
}
\label{o2temp}
\end{figure*}
\begin{figure*}
\figurenum{6}
\plotfiddle{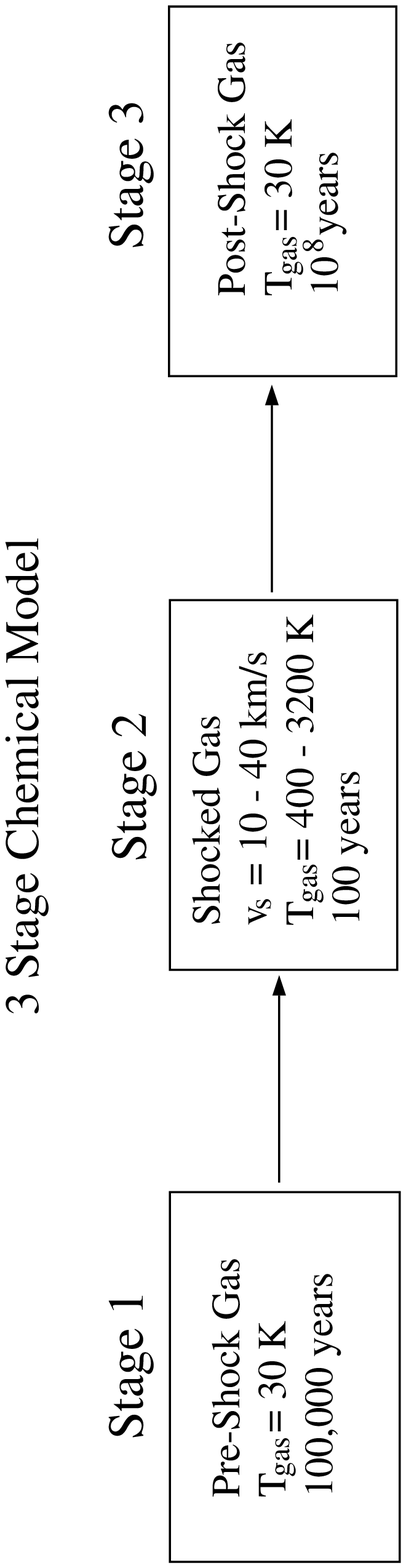}{2.5in}{270}{65}{65}{-260}{300}
\caption{
Schematic demonstrating the physical evolution of the three-stage chemical model of 
pre-shock, shocked, and post-shock gas.  In all stages the density, visual extinction,
and UV field enhancement are kept constant
at $n_{H_2} = 10^5$ cm$^{-3}$, $A_{V} = 10$ mag, $G_{\rm o} = 1$. 
}
\label{3stage}
\end{figure*}

\subsection{Three-Stage Chemical Model}

We envisage that the chemical evolution of gas in star-forming molecular 
clouds that have been subjected to non-dissociative shocks as one with 
several potentially episodic
evolutionary stages.  The first stage of evolution is simply quiescent chemical
and physical evolution.  The gas starts its chemical evolution at low density
with mostly atomic constituents and proceeds towards higher density with
an increasing molecular composition.
At some time the cloud undergoes a period of star formation
and the protostars alter the local environment through radiation,
winds, and outflows.   The second stage of evolution occurs when a parcel of molecular
gas is processed by a shock.  After the passage of a shock, the gas
rapidly cools, and the pre-shock time-dependent chemistry begins to be re-established. 
Thus the post-shock chemistry represents the third evolutionary stage.   
For a given parcel of gas, these three stages can be repeated, provided that
star formation is still active in nearby regions (c.f. Charnley et al. 1988a,b).

To account for this picture, we have constructed a three-stage chemical model, shown
schematically in Figure 6. In Stage (1)
we use a pseudo-time dependent calculation of the chemical evolution in which
the gas is initially atomic and
the physical conditions for quiescent gas are those described earlier.  
This evolution continues with constant physical properties
until $t = 10^{5}$ years when the gas
undergoes a 20 km s$^{-1}$ non-dissociative shock, which heats the gas and raises 
the gas temperature.
At the completion of Stage (1), Stage (2) is started by re-initializing
the chemical evolution and
raising the gas temperature to the maximum value expected for a 20 km s$^{-1}$
shock ($T_{gas} \approx 1000$ K; see Figure 1).   In this fashion, the chemical
abundances after
10$^5$ years of quiescent evolution are the initial conditions for the second stage.
Based on the discussion in the
previous section, we allow the evolution to continue at high temperature for 
100 years.   After 100 years, Stage (3) is begun and
the chemistry is re-initiated with the abundances of 100 years of evolution
at elevated temperatures providing the initial abundances. 
For the third stage we use the original gas temperature ($T_{gas} = 30$ K) 
and allow the chemistry to re-equilibrate.  
Through each of these stages, pre-shock (Stage 1), shock (Stage 2), and post-shock
(Stage 3), only the temperature is altered in step-wise fashion, 
the other physical properties (\nhtwo , $G_{\rm o}$, $A_V$) remain constant.  
The step-wise time evolution of temperature is a simple approximation of the 
shock temperature profile shown in Figure 2.  
As a check on these results
we have also run a more complex multi-stage model using this temperature profile
and we find that the results which we present in the following paragraphs
are unaltered.

\subsubsection{Pure Gas-Phase Chemistry}

We first examine the three-stage chemical model using only the gas-phase chemical
network.  Figure 7 presents the results from a three-stage chemical model 
of a 20 km s$^{-1}$ shock impacting on quiescent gas.  In this figure the top
panel presents the pre-shock chemical evolution 
(Stage 1), the middle panel the evolution in shocked gas (Stage 2), and the bottom 
panel the chemistry of the post-shock gas layer (Stage 3).  {\em Note that the temporal range
is different for each panel.}

\begin{figure*}
\figurenum{7}
\plotfiddle{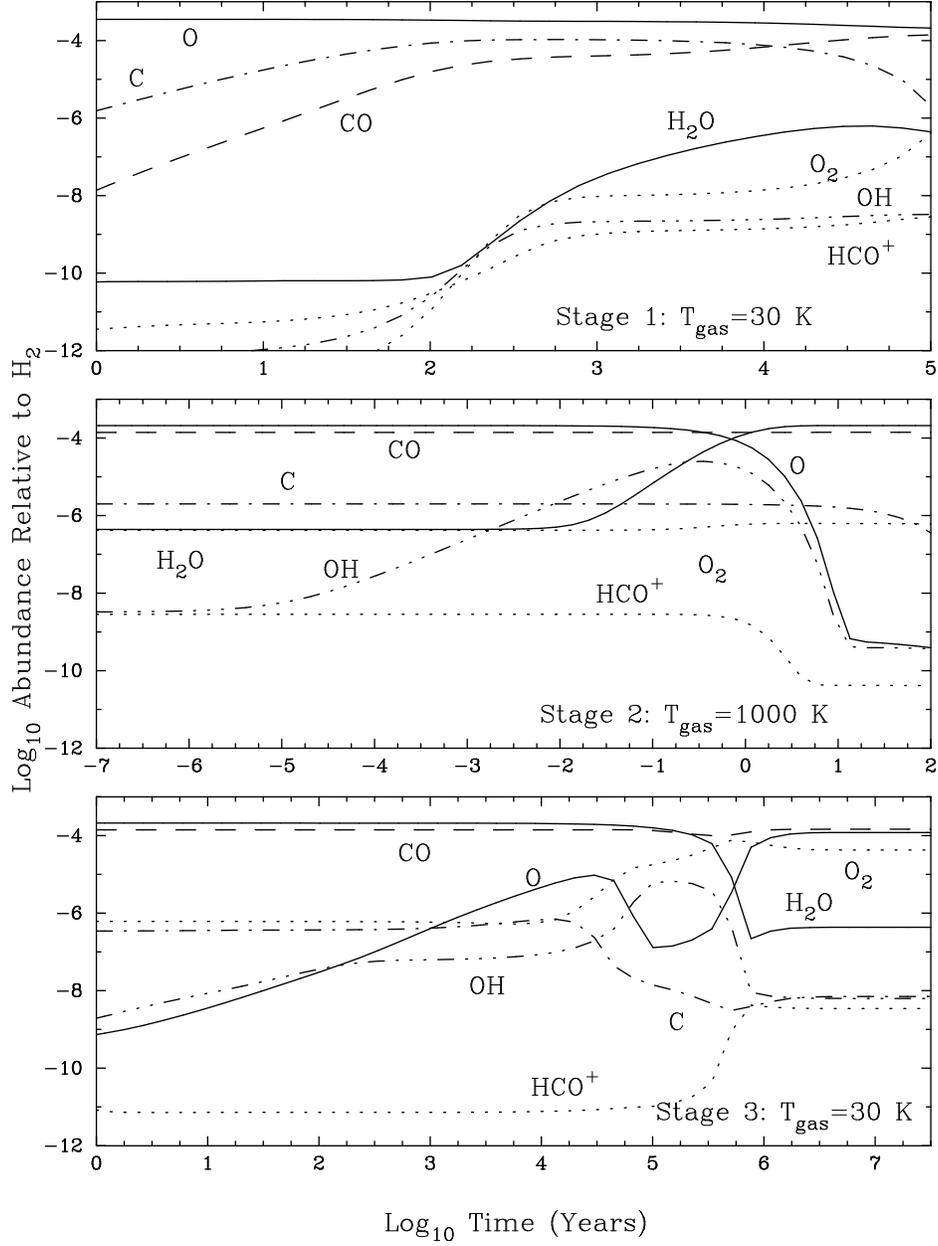}{7in}{0}{70}{70}{-200}{0}
\caption{
Evolution of chemical abundances relative to H$_2$ in the pure gas phase three-stage model
as a function of time for simple carbon- and 
oxygen-bearing molecules (C, CO, O, OH, O$_2$, H$_2$O, HCO$^{+}$).
Stage 1 (top panel) represents the pre-shock stage with $T_{gas} = 30$ K, 
Stage 2 (middle panel) is the shock stage
with a higher gas temperature ($T_{gas} = 1000$ K), and Stage 3 (bottom panel) 
the post-shock stage with $T_{gas} = 30$ K.  Note that the time axis for 
each panel covers a different range and shows the full evolution of
chemical abundances from one stage to the next.
}
\label{3stage_m1}
\end{figure*}

Examining Figure 7 we observe that the chemical abundances in Stage (1) 
are quite similar to that observed 
in Figure 3; carbon is slowly processed into CO and the abundances \HtwoO\ 
and \Otwo\ steadily increase.   In Stage (2) the temperature is raised, and
the \HtwoO\ abundance quickly rises to incorporate all of the oxygen that is
not locked into CO.  The formation of \HtwoO\ from O is preceded by formation of OH,
which accounts for the sharp rise and decline in the hydroxyl abundance.
However, even at 1000 K, the abundances of \Otwo\ and CO are unchanged.
In the post-shock evolution (Stage 3), the abundances of most species do
not change for $\sim 10^4$ years.  During this time there is a noticeable
increase in the abundance of atomic oxygen, suggesting that even for 
times $< 10^{5}$ years there is some destruction of \HtwoO .  But, for
these early times,
the abundance of atomic oxygen is more than an order of magnitude below that of
\HtwoO .  After $\sim 4 \times 10^5$ years of low-temperature evolution the chemistry 
dramatically changes as the abundance of
water rapidly drops and the ``normal'' chemistry, with CO, O, and
\Otwo\ as the dominant oxygen reservoirs, is re-asserted.  
At this time the abundance of CO also exhibits a small drop followed by an
immediate increase, which is due to 
CO reacting with OH (producing CO$_2$).   This reaction has been measured
at low temperatures in the laboratory (\cite{FSS91}).    

\begin{figure*}
\figurenum{8}
\plotfiddle{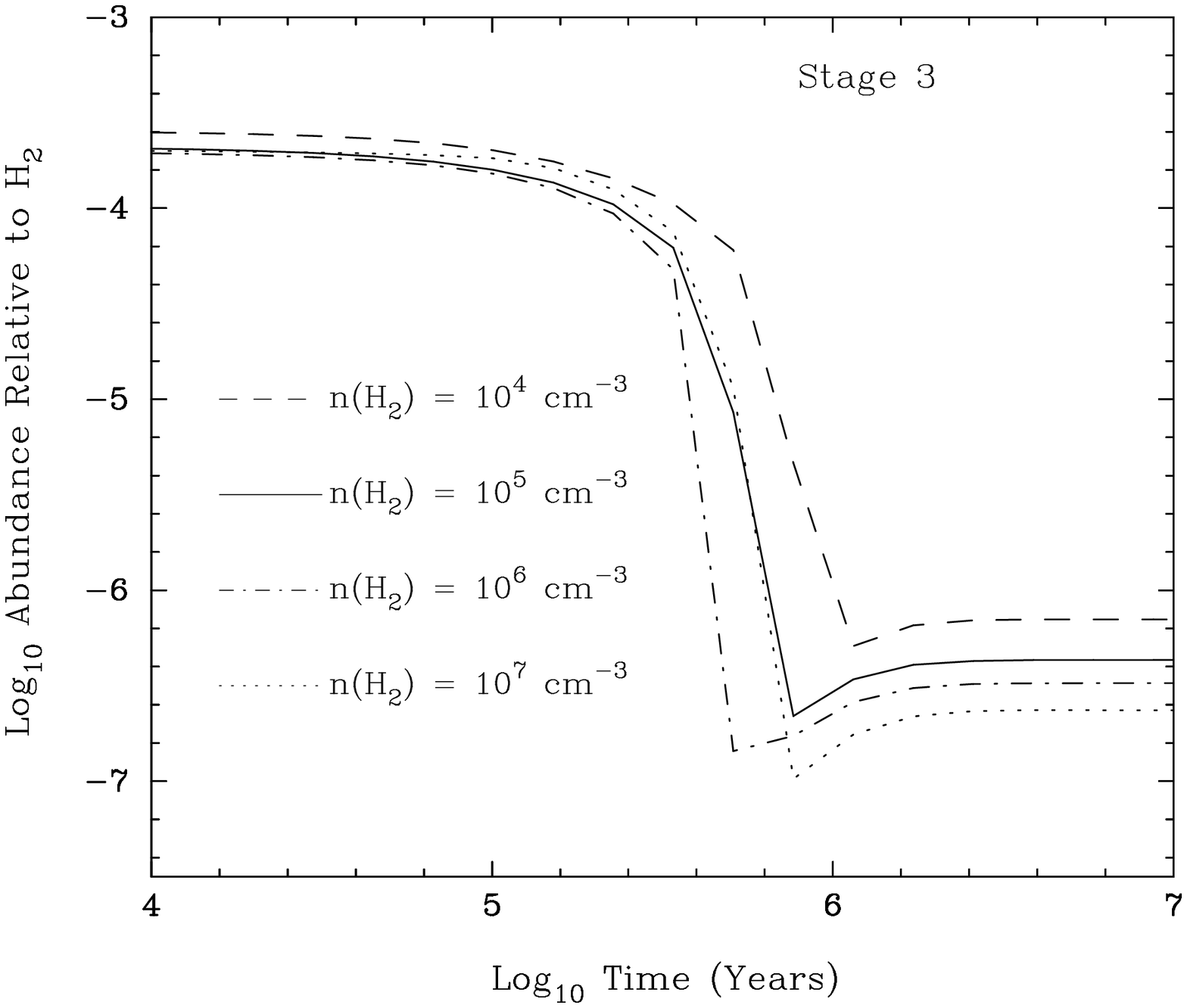}{2.5in}{0}{65}{65}{-200}{-175}
\caption{
Evolution of H$_2$O abundance relative to H$_2$ in Stage 3 shown for $n_{H_2} = 10^{4}$, 10$^{5}$,
10$^{6}$, and 10$^7$ cm$^{-3}$.
}
\label{h2oden}
\end{figure*}

In Figure 7 the quiescent stage is allowed to evolve for 10$^{5}$ years,
if this stage is allowed to evolve for 10$^{6}$ years,
or even 10$^{7}$ years, the abundances of a few selected species in the subsequent
stages would be altered.
Examining the full quiescent evolution shown in Figure 3, if the first stage evolves
beyond 10$^5$ years,
the abundance of atomic carbon is lowered, while 
molecular oxygen will be more abundant.  If the shock is not strong enough to 
destroy \Otwo , the larger \Otwo\ abundance would trap oxygen in molecular form
and produce a lower \HtwoO\ abundance in the shock and post-shock stages. 

For \nhtwo\ $= 10^5$ \cc , the timescale for the re-assertion of low-temperature
chemistry, $\tau_{ra}$, is $\sim 4 - 7 \times 10^5$ years.   In Figure 8 we
examine the dependence of $\tau_{ra}$ on density for three-stage models run with
\nhtwo\ $= 10^4, 10^5, 10^6, 10^7$ \cc .  In this figure we present only the
water abundance in the post-shock stage.  Even though chemical interactions
occur with greater frequency at higher density,  the chemical relaxation timescale 
does not depend on the density.   This is a result 
of cosmic-ray driven chemistry.   The principle mechanism for re-distribution
of water back to other oxygen-bearing species is through the following two reactions:

\begin{equation}
\eqnum{R5}
\rm{H_2O + H_3^+ \rightarrow H_3O^+ + H_2.}
\end{equation} 

\begin{equation}
\eqnum{R6}
\rm{H_2O + HCO^+ \rightarrow H_3O^+ + CO.}
\end{equation} 

\noindent The first channel is important at early times, but the second channel
dominates at late times. 
Because of the large abundance of water, the abundances of \HCOp\ and \Hthreep\
decrease in the shock phase.  This effect on \HCOp\ is shown in Figure 7.  However, in the third stage
at $t > 10^5$ years, the abundance of \HCOp\ and \Hthreep\ begin to increase and
water begins to be efficiently destroyed through the above reactions.
Subsequently, the destruction of \HCOp\ and \Hthreep\ decreases and   
the process begins to feed itself and the water abundance rapidly drops.  

The lack of a dependence of $\tau_{ra}$ on density
is also an important clue because, while the fractional abundance of \Hthreep\
depends on the density, the space density of \Hthreep\ ($x(\rm{H_3^+})$\nhtwo )
is constant with density (\cite{LDS87}).
Because $\tau_{ra}$ is tied to the abundance of \Hthreep , it
will depend on the cosmic-ray ionization rate.  In these calculations we have used 
an ionization rate of $\zeta_{H_{2}} = 10^{17}$ s$^{-1}$;
for an ionization rate a factor of 5 higher, the timescale decreases
approximately by a factor of 5.

\begin{figure*}
\figurenum{9}
\plotfiddle{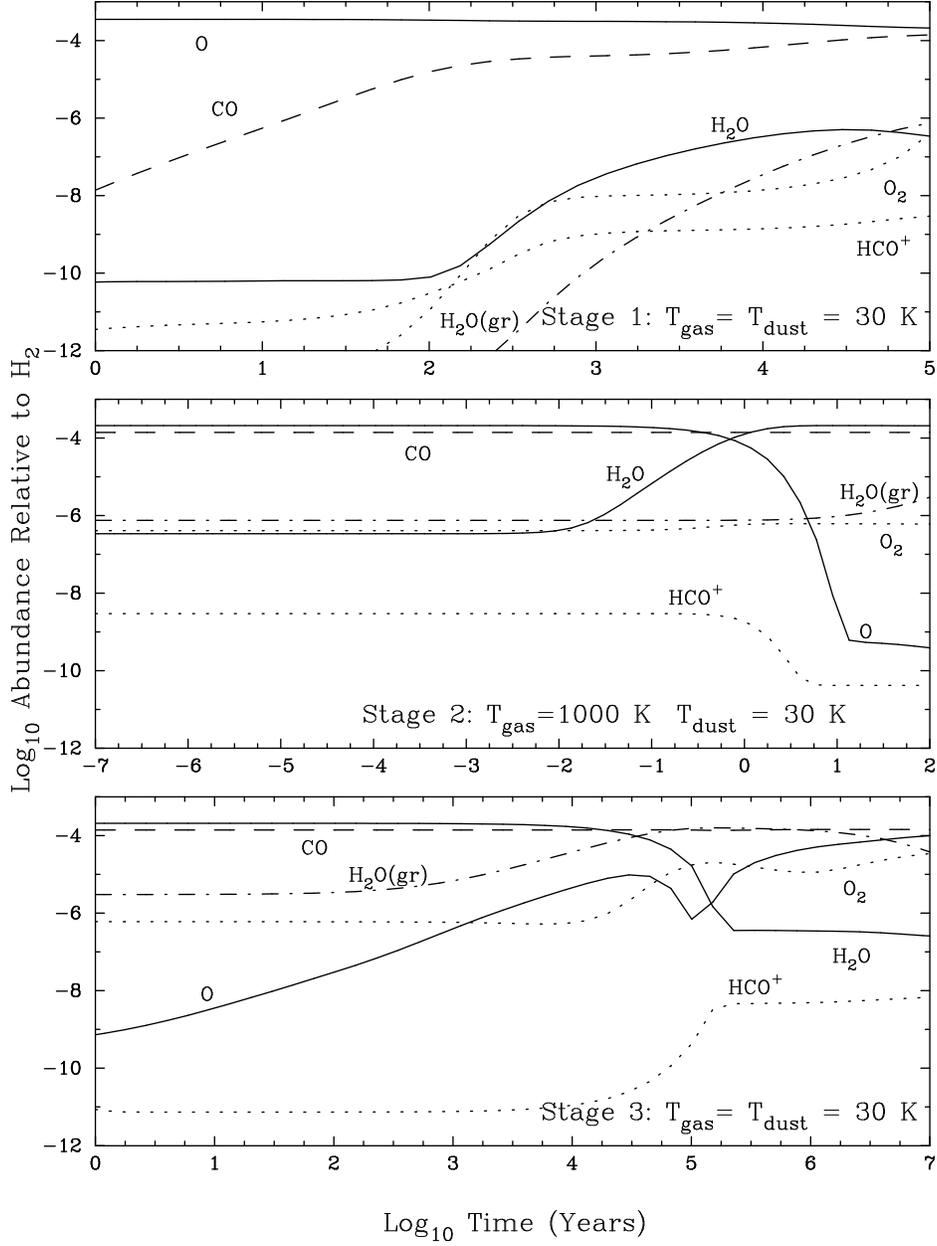}{7in}{0}{70}{70}{-200}{0}
\caption{
Evolution of chemical abundances in the gas-grain three-stage model
as a function of time for simple carbon- and 
oxygen-bearing molecules (C, CO, O, O$_2$, H$_2$O, HCO$^{+}$).
All abundances are relative to H$_2$.
This model includes molecular depletion and desorption from grain surfaces and
the abundance of H$_2$O on the grain surface (H$_2$O(gr)) is presented.
Stage 1 (top panel) represents the pre-shock stage with $T_{gas} = 30$ K, 
Stage 2 (middle panel) is the shock stage
with a higher gas temperature ($T_{gas} = 1000$ K), and Stage 3 (bottom panel) 
the post-shock stage with $T_{gas} = 30$ K.  
Note that the time axis for 
each panel covers a different range and shows the full evolution of
chemical abundances from one stage to the next.
}
\label{3stage_m1gr}

\end{figure*}
\subsubsection{Gas-Phase Chemistry Including Gas-Grain Interactions}

\begin{figure*}
\figurenum{10}
\plotfiddle{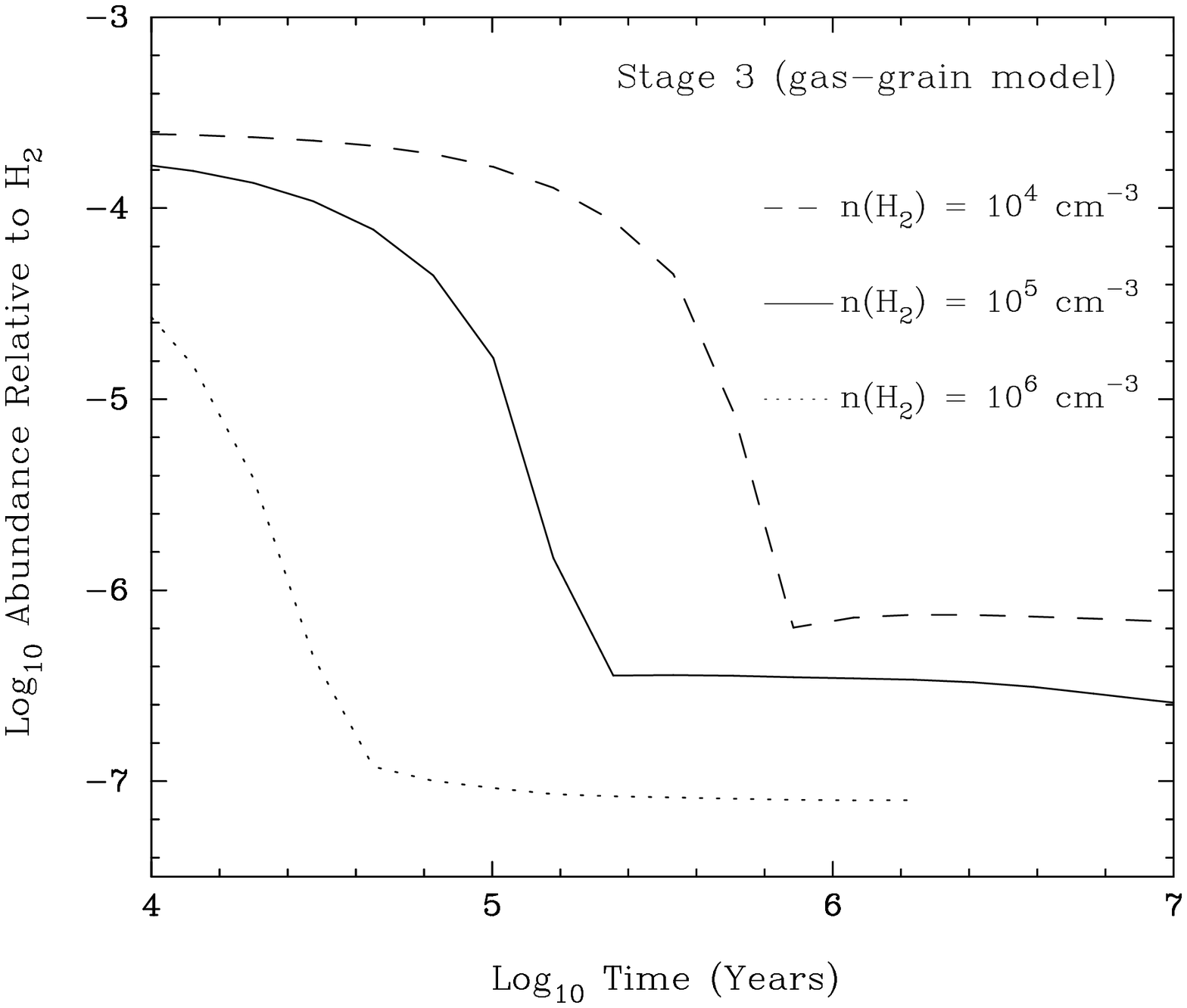}{2.5in}{0}{65}{65}{-200}{-180}
\caption{
Evolution of H$_2$O abundance in Stage 3 from the gas-grain model shown for $n_{H_2} = 10^{4}$, 10$^{5}$,
and 10$^{6}$ cm$^{-3}$.   Abundances are relative to H$_2$.
}
\label{h2odengr}
\end{figure*}

Figure 9 presents the three-stage model run for \nhtwo $= 10^5$ \cc, including
grain surface molecular depletion and desorption.  
For these models we have set
the dust temperature equal to the gas temperature $T_{dust} = T_{gas} = 30$ K.
When the gas temperature is raised in Stage 2, the dust temperature
is kept constant (c.f. \cite{DRD83}).  
The dominant desorption mechanism
in this model is thermal evaporation from 30 K dust grains.  The desorption rate
is $k(H_2O)_{evap} = 7.2 \times 10^{-15}$ s$^{-1}$ (see \cite{BLG95}). 
In this model we allow for the depletion of molecular species onto grain surfaces
and account for some desorption, but we do not allow species to react on the grain 
surfaces.  For water-ice this assumption is not unreasonable because \HtwoO\
is a stable endpoint in models of surface chemistry (c.f. Tielens and Hagen 1982)
and the majority of gas phase
\HtwoO\ depletion will occur in the post-shock layer when the water abundance has been
raised by the shock.
In addition, icy grain mantles may be removed inside shocks (through
sputtering) such that 
all of the oxygen will eventually exist in the gas phase in the form of water.
Thus, water created by the high--temperature shock will deplete back onto
the mantle directly from the gas phase and 
will not be formed 
through grain surface reactions. 
For the three-stage model we do not account for any sputtering of the mantle 
inside the shock, but we will account for sputtering in the more general model
presented later. 

The quiescent evolution shown in the top panel of Figure 9 is not appreciably
different from that seen in the pure gas-phase case.  However, there is slow but steady
accretion of water from the gas phase, such that by $t = 10^{5}$ years 
water is more abundant on the grain surface than in the gas phase.  In contrast,
due to lower grain surface binding energies, and therefore enhanced evaporation rates,
both \Otwo\ and CO remain in the gas phase.  In Stage 2 the 
evolution between gas-phase and gas-grain models is almost identical; 
all of the oxygen not in CO, or on grains, is processed into \HtwoO .
In Stage 3 we see the largest differences.  Here the abundance of
gas-phase water exhibits a sharp decline at $t \sim 10^5$ years, which is
less than $\tau_{ra}$ inferred from pure gas-phase chemistry.  
The decrease in gas-phase water abundance
is the result of depletion onto grains, which has a timescale of
$\sim 10^5$ years for this density.   
As a result, the abundance of water on grains at these times
becomes quite large $\sim 10^{-4}$.
At later times ($t > 10^{6}$ years), the abundance of \HtwoO\ on the
grain surface decreases as the water 
molecules in the grain mantles begin to escape via thermal evaporation.
Because CO and \Otwo\ do not deplete, the oxygen chemistry is maintained
in the gas phase and the overall gas-phase water abundance is only a factor of 2 or
3 below that of the pure gas-phase case.  We note that similar chemical evolution
will be observed for lower dust temperatures because both CO and \Otwo\ will not deplete
from the gas phase (due to cosmic-ray-induced heating), while the more tightly bound
\HtwoO\ molecules will deplete from the gas phase (see \cite{BL97}).

To examine the dependence of post-shock gas-grain chemistry on density,
in Figure 10 we
present Stage 3 for models run at three different densities, 
\nhtwo\ $= 10^4, 10^5, 10^6$ \cc .
This figure demonstrates that the inclusion of grains places a density dependence on
$\tau_{ra}$, which is a reflection
of the depletion timescale for water. For \nhtwo\ $= 10^4$ \cc, the depletion
timescale is equal to $\tau_{ra}$ while for higher densities the depletion
timescale is less than the gas-phase relaxation timescale.
Lastly, due to the large abundance of water trapped on grains,
the equilibrium chemistry for gas-grain models is never truly re-asserted
since the chemistry does not reach equilibrium.

\subsection{Monte-Carlo Model of Molecular Cloud Evolution}

The results presented in Section 4.3 suggest that if the frequency of shock
passages is similar to $\tau_{ra}$, then the water abundance should be quite
large ($\sim 10^{-4}$).  
In contrast, if the shock timescale is $\gg \tau_{ra}$, then the
water abundance should be close to the lower values expected 
in quiescent gas ($\sim 10^{-7}$).
In this sense, the abundance of \HtwoO\ is a measure of the physical shock history
within a cloud -- more so than the dynamical information contained
in line profiles, which can be short-lived compared to the lifetime
of the chemical legacy.  Likewise, because \Otwo\ is destroyed in very
strong shocks, the abundance of \Otwo\ may also chart the history
of such shocks.

\begin{figure*}
\figurenum{11}
\plotfiddle{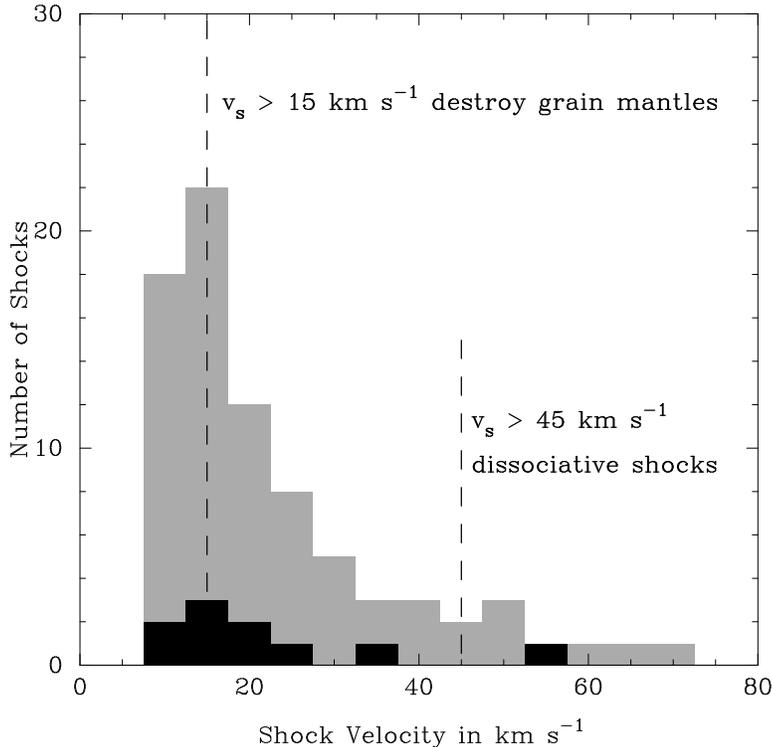}{2.5in}{0}{65}{65}{-220}{-190}
\caption{
Number of shocks in 5 km s$^{-1}$ bins from Monte-Carlo model run for $\sim 10,000$ 
cycles.  Two models are shown, $\tau_s = 10^6$ (shaded histogram) and 10$^7$ years
(filled histogram).
}
\label{shockspec}
\end{figure*}
\begin{figure*}
\figurenum{12}
\plotfiddle{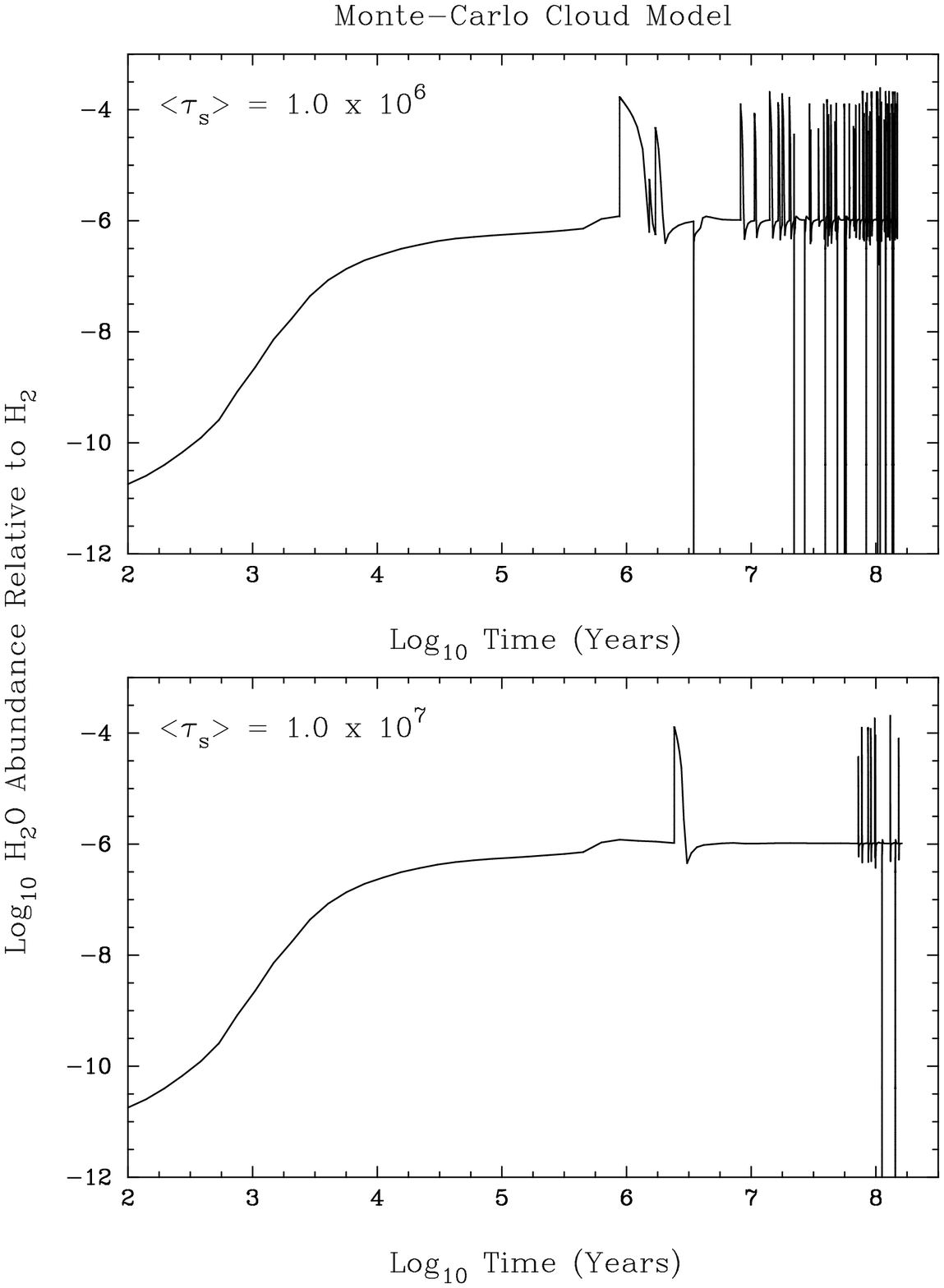}{7in}{0}{70}{70}{-230}{-40}
\caption{
Abundance of H$_2$O relative to H$_2$ 
shown as a function of time for the Monte-Carlo cloud model.
The top panel presents the chemical evolution for $\tau_s = 10^6$ years and the bottom
panel for $\tau_s = 10^7$ years.  The vertical lines extending to abundances below 10$^{-12}$
are dissociative shocks.
}
\label{mcmodel}
\end{figure*}
To examine the utility of using the abundance of \HtwoO\ and \Otwo\ 
to trace the shock history of molecular gas we have created a more general
model of shocks in clouds using a Monte-Carlo simulation.   
%First, it
%must be recognized that
%over the lifetime of a cloud gas might be processed in a shock several times before
%it collapses to form a star, or is dispersed and the molecules dissociated by the
%radiation field.  This type of cyclic model is supported by observations of
%B5 (Goldsmith, Langer, \& Wilson 1986) and provides the basis for the chemical
%limit models of Charnley et al. (1988a,b).  
To construct the Monte-Carlo cloud model we use the 
10 km s$^{-1}$ shock timescales determined in Section 3 for massive star-forming
clouds and the chemical-dynamical model discussed in Section 2.
In this model we allow the chemistry
to evolve in steps of 10$^{4}$ years.  
If the shock timescale is $<\tau_{s}> = 10^{6}$ years,
then during each time interval there is a 1 percent chance that the gas will experience 
a shock.  
This model is similar to the three-stage model,
except that the time evolution of each stage is 10$^{4}$ years.  As before,
when a shock passes (determined by the generated random number), 
it is assumed that in all cases the period of elevated
temperatures is 100 years; for all other times it is assumed that
$T_{gas} = 30$ K.   
As a consequence of the momentum driven winds discussed in Section 3, we
assume that the frequency of shocks faster than a given velocity
is inversely proportional to that velocity.  Thus 20 km s$^{-1}$ shocks
occur with a frequency that is a factor of two below that for shocks faster than
10 km s$^{-1}$ shocks.   
The correspondence between the shock strength (velocity) and post-shock temperature
is presented in Figure 1.  
If $v_s > 45$ km s$^{-1}$ we account for the 
possibility of {\em dissociative} shocks and destroy all molecular abundances.
The next cycle is then begun with the atomic initial abundances. 
The entire model is continued for a total of $\sim 10^{8}$ years or $\sim 10,000$ cycles.

The computations required to run thousands of cycles is too time consuming to
use the full UMIST database.  We therefore constructed a smaller network of
26 species and 145 reactions.  This small network
contained all of the important high- and low-temperature pathways to
form carbon and oxygen species and was tested against a sample run that made
use of the entire UMIST database, with no significant differences found.
%We again separately examine the pure gas-phase evolution and the chemistry
%including gas-grain interactions. 

Finally, models of shocked molecular gas have found that species can be
desorbed from the grain mantle, or even the grain core, in shocks through sputtering
or grain-grain collisions (c.f. \cite{DRD83}; \cite{SWPdFF97}; \cite{CHH97}).  
The amount of gas removed depends on the
shock velocity and  differs between various models.
Based on the results of Caselli et al. (1997) we assume that shocks with
$v_s \ge 15$ km s$^{-1}$ will entirely remove the grain mantle.  In these models
we do not explicitly include a sputtering term, but rather simulate these
effects by raising the dust temperature to 100 K.  This will increase the rate
of thermal evaporation and rapidly desorb mantle species. 
 
\subsubsection{Pure Gas-Phase Chemistry}

In Figure 11 we show the number of shocks in 5 km s$^{-1}$ bins that are generated
by the Monte-Carlo cloud simulation with $<\tau_s> = 10^6$ years (shaded histogram)
and 10$^{7}$ years (filled histogram).   From the discussion of shock timescales in
Section 3, $<\tau_s> = 10^6$ years is a value appropriate for regions of massive
cluster formation (e.g., OMC-1), while  $<\tau_s> = 10^7$ years is inferred for
general
GMC core material (e.g., Mon R2 or L1641).  In both cases, the spectrum of shocks 
is heavily weighted towards lower velocities, but there
are several fast dissociative shocks, even for $<\tau_s> = 10^7$ years.  
There are a few shocks with
velocities higher than 80 km s$^{-1}$, but they are not included in this figure. 

Figure 12 shows the full time dependence of the water abundance 
from the gas-phase Monte-Carlo model with
$<\tau_s> = 10^6$ years (top panel) and 10$^{7}$ years (bottom panel)
using the shock spectrums shown in the previous figure.
For $<\tau_s> = 10^6$ years, the water abundance increases 
for the first 10$^5$ years and at $t \sim
10^6$ years sharply rises as the gas is shock-heated. 
The abundance decreases in the post-shock gas and goes through a series of 
successive shock heating and cooling events.  
The vertical lines that extend below $x(\rm{H_2O}) = 10^{-12}$ are
the result of dissociative shocks.  
Overall, the water abundance varies between $x(\rm{H_2O}) \sim 10^{-6}$ and $10^{-4}$,  
with an average abundance of $\sim 10^{-5}$.
For a longer shock timescale (bottom panel), the water abundances show similar
variations, but the effects are not as dramatic.

\begin{figure*}
\figurenum{13}
\plotfiddle{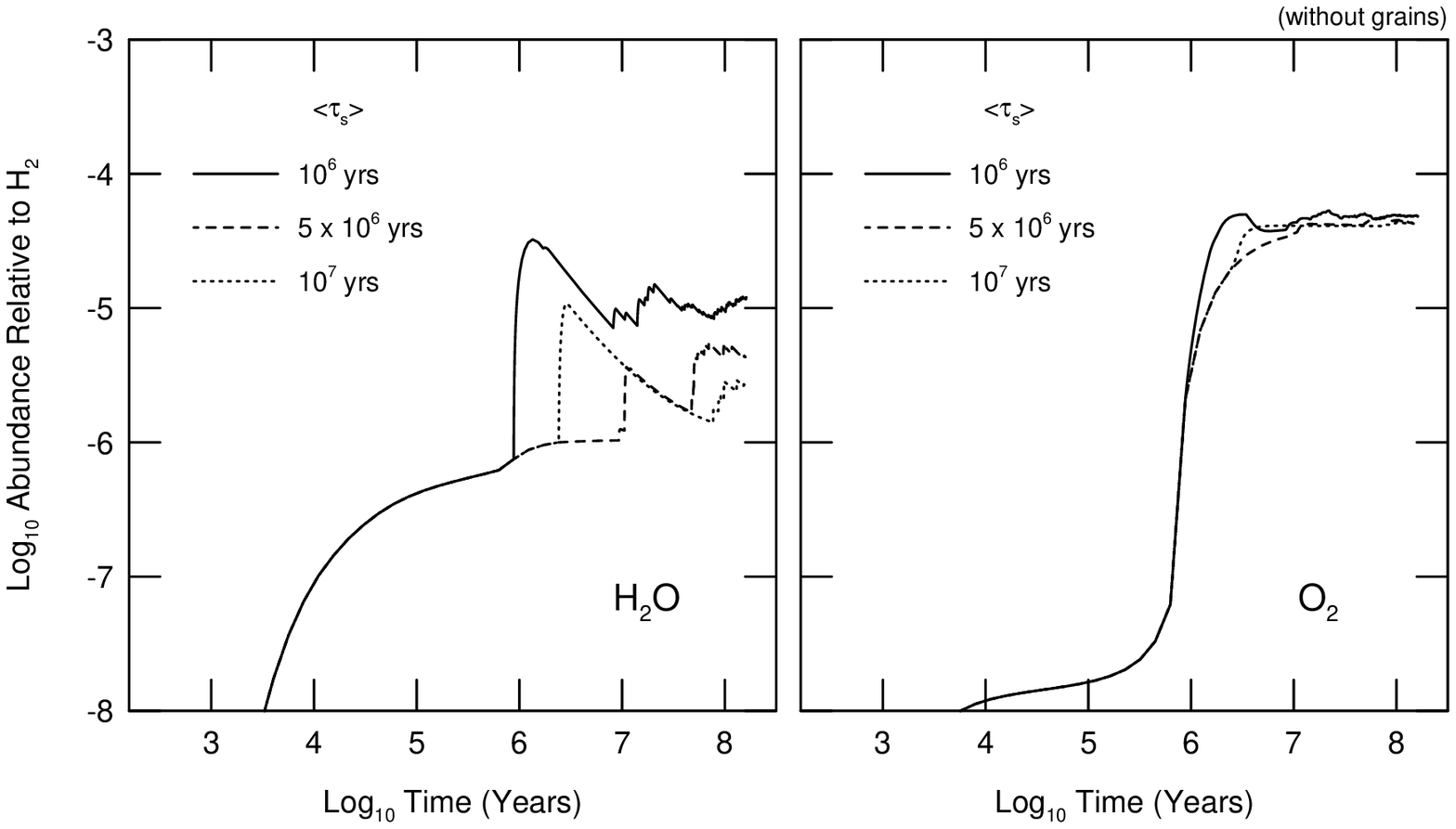}{2.5in}{0}{80}{80}{-240}{-50}
\caption{Time average abundances of H$_2$O and O$_2$ from Monte-Carlo cloud models run
with variable shock timescales.  This model is for pure gas-phase chemistry 
at $n_{H_2} = 10^5$ cm$^{-3}$ and all
abundances are calculated as relative to H$_2$.  The time-averaged abundances are calculated
as a running average, weighted by the amount of time the abundance remains at a given
amount.
}
\label{timeavg}
\end{figure*}

\begin{figure*}
\figurenum{14}
\plotfiddle{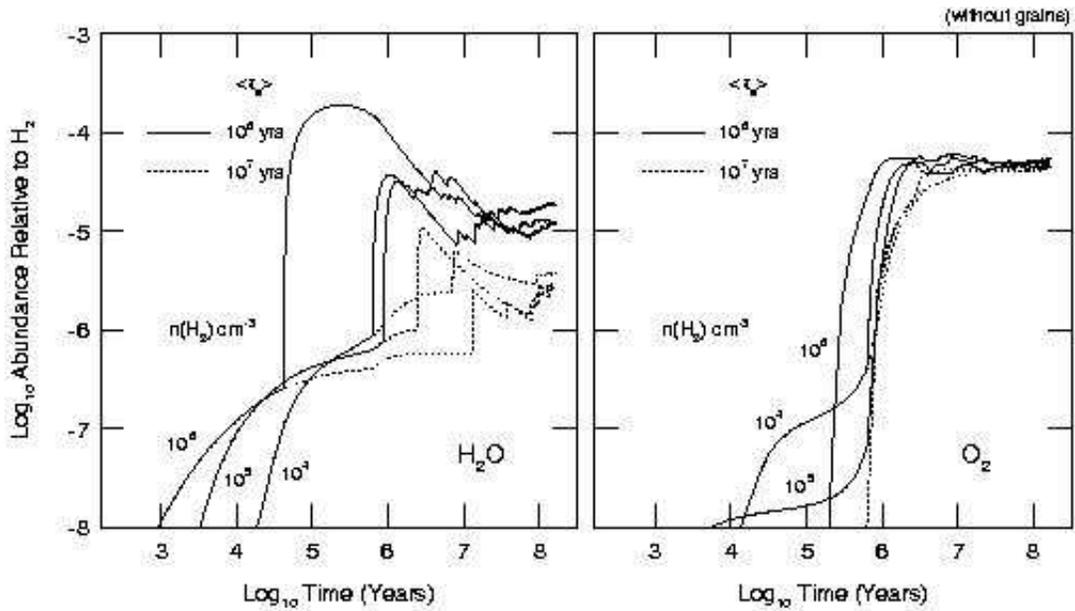}{2.5in}{0}{80}{80}{-240}{-200}
\caption{
Time averaged H$_2$O and O$_2$ abundances (relative to H$_2$) from Monte-Carlo models
run with two separate shock timescales and three different densities.
}
\label{timeavgden}
\end{figure*}

\begin{figure*}
\figurenum{15}
\plotfiddle{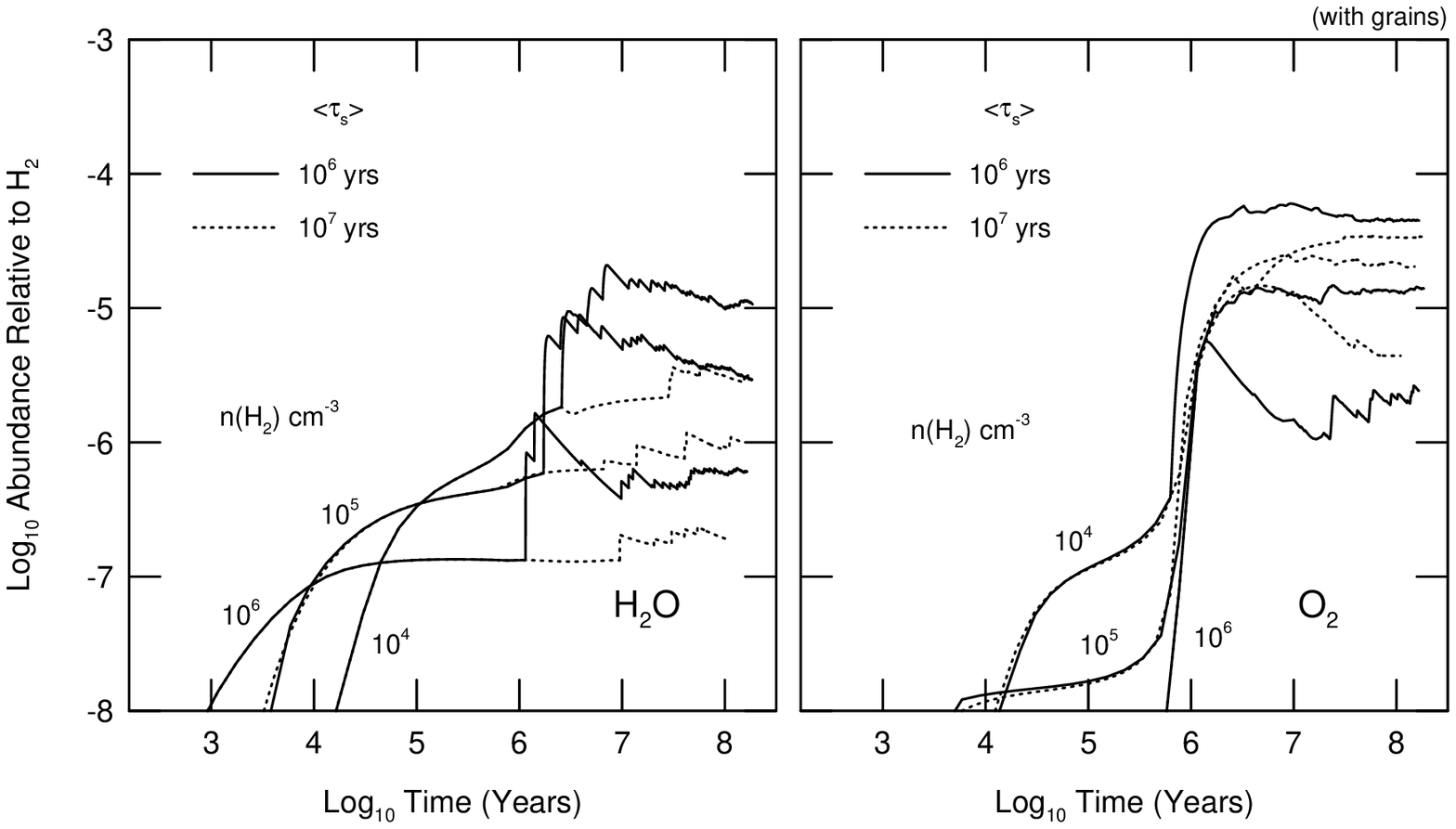}{2.5in}{0}{80}{80}{-240}{-50}
\caption{Same as in Figure 13, but for gas-grain Monte-Carlo model.
}
\label{timeavggr}
\end{figure*}
We have computed the time-averaged \HtwoO\ and \Otwo\ abundances in the 
model presented in Figure 12 and these results are shown in Figure 13.  
To create this plot we compute a running average as a function of time,
weighting the abundance by the length of time it remains at a given value.
In this figure we also present the time-averages for 
another Monte-Carlo model run with $<\tau_s> = 5 \times
10^6$ years.  For $<\tau_{s}> = 10^{6}$ years, 
the time-averaged water abundance converges 
to $\sim 10^{-5}$.  For other, longer shock timescales, the water abundance 
takes longer to converge to a single value, 
but the overall amount of water is clearly dependent on the 
shock timescale.   These effects can be quite large for $<\tau_s>$ greater
than 10$^7$ years; 
the average water abundance will be at least a factor of five below
the value for $<\tau_s> = 10^6$ years. 
It is also notable that even when the shock timescale is $10^7$ years,
the frequency of shocks is high enough that the time-averaged water abundance is  
greater than that expected for steady-state chemistry in quiescent gas (compare with
Figure 1 where $x(\rm{H_2O}) \sim 6 \times 10^{-7}$).
In contrast, because the frequency of shocks with velocities high
enough to destroy \Otwo\ within the cooling time is low,
the abundance of \Otwo\ has a negligible dependence on the shock timescale.

Figure 14 presents the time-averaged abundances for
models run at $<\tau_s> = 10^6$ and $10^7$ years for three different densities.
The results shown in this figure confirm the earlier finding that for pure
gas-phase chemistry and a constant cosmic-ray ionization
rate, the chemical relaxation timescale has little dependence on the 
density.  Similarly, the time-averaged \HtwoO\ abundance is not dependent
on density.  While the relaxation timescale does not depend on the density it 
does have a dependence on the cosmic-ray ionization rate (see discussion in Section
4.3.1).  The relaxation timescale is inversely proportional to the ionization rate,
such that a Monte Carlo model with a shock timescale $\tau_s$ and cosmic 
ray ionization rate  of $\zeta_{H_{2}} = 10^{17}$ s$^{-1}$ would roughly apply for  
a model with a shock timescale $<\tau_s>' = <\tau_s> / (\zeta_{H_2}'/10^{-17} s^{-1}$), 
where $\zeta_{H_2}'$ is a different cosmic ray ionization rate.

\subsubsection{Gas-Phase Chemistry Including Gas-Grain Interactions}

The time-averaged abundances calculated in the Monte-Carlo model, which
include interactions between the gas phase
with grain surfaces, are shown in Figure 15.  These results are presented for
three densities, \nhtwo\ $= 10^4$, $10^5$, $10^6$ \cc\ and two shock timescales
$<\tau_s> = 10^6$ and 10$^7$ years.   As before, 
species are fully evaporated from the grain mantles if shock 
velocities exceed 15 km s$^{-1}$.  The inclusion of depletion onto grains
introduces a density dependence into the overall water abundance.
Although the \HtwoO\ abundance does converge to a single time-averaged value, this
value does have a dependence on the shock timescale.  Therefore, to compare
these results with observations, prior information about
the density is required in order to differentiate between the various solutions.
For molecular oxygen,
the inclusion of gas-grain interactions introduces a dependence on the average abundance 
with density (because of the increasing depletion rate with increasing density).  
However, the time-averaged \Otwo\ abundance is not as strongly dependent on the 
shock timescale.

\section{Discussion}

\subsection{Comparison with Observations}

Our results demonstrate that the {\em time-averaged} abundance of \HtwoO\
is a sensitive function of the rate at which molecular gas is subjected to shocks.
In order to properly interpret these results it is useful to consider 
how the time-averaged abundances relate to observations of water
in a giant molecular cloud (GMC).   The gas inside a star-forming
cloud can roughly be divided into three categories:  (1) quiescent
material that has been relatively unaffected by current or previous
epochs of star formation; (2) gas that is being physically and/or
chemically affected by current star formation; and, (3) as suggested here,
gas that has been affected by a 
previous generation of star formation and is in the process of chemically 
and dynamically evolving back to a 
quiescent state.  Observations of \HtwoO\ toward molecular material
associated with each of these categories should find abundances
varying from as low as $x(\rm{H_2O}) \sim 10^{-6}$ in (1), and ranging upwards to
$\le 10^{-4}$ in (2) and (3).   {\em A single pointed detection of H$_2$O 
emission in a GMC can therefore be
represented by a single stage of the three-stage or Monte-Carlo models
(quiescent, shock, post-shock).  Because star formation will be spread
throughout a cloud or core, the time-averaged
abundance therefore refers to water abundances averaged over an area that
contains both quiescent material and any associated gas currently being
affected by local star formation.
Thus the average abundance over a cloud (cloud-average) is comparable to the
time-averaged abundances in the Monte-Carlo cloud model.}  This average 
abundance might not necessarily apply to an entire GMC complex
(e.g. Orion, Gem OB1), which can extend for several square degrees on
the sky, but may apply to several dense cores within a
single cloud. 
Regions with high star-formation rates, such as OMC-1 which is associated
with the Trapezium cluster (see Section 3), 
might be expected to have a higher probability for strong shocks,  and therefore a high
cloud-averaged water abundance.  Cores
with lower star-formation rates, such as L1641 also in Orion,
would have low cloud-averaged water abundances.  

{\em To test our model's ability to constrain the history of molecular
clouds it is important to obtain maps of the water emission
over large spatial scales in GMC cores.}
Because of the strong absorption by atmospheric water vapor,
observations of \HtwoO\ in the ISM are extremely difficult.
As a result, the small number of detections of water in molecular
clouds have typically been taken towards single lines of sight 
mostly containing luminous protostars.
The very convincing detections of water by ISO provide the
greatest evidence for high water abundances in hot gas, with $x(\rm{H_2O}) 
\sim 1 - 6 \times 10^{-5}$
inferred towards hot stars (\cite{Helmich_etal96}; \cite{vDH96}) and HH54
(\cite{Liseau_etal96}).  Towards Sgr B2 Zmuidzinas et al. (1994) observed \HtwoeiO\ in
absorption with the Kuiper Airbourne Observatory (KAO),
while Neufeld et al. (1997) observed \HtwoO\ in emission using ISO.
Combining these observations with previous ground based detections
Neufeld et al. (1997) estimate
an abundance of  $x(\rm{H_2O}) = 3.3 \times 10^{-7}$
for the cooler outer parts of Sgr B2 and  $x(\rm{H_2O}) \sim 5 \times 10^{-6}$ in the
hot core.   
These observations are important not only because \HtwoO\ was
detected in dense gas, but also because there appears to be a range of
water abundances.  However, it is difficult to discern 
whether the enhanced water abundances are the result of high-temperature chemistry
that occurs behind shocks, high-temperature chemistry appropriate to gas near
young stars (c.f. Doty \& Neufeld 1997),
or evaporation of grain mantle species (see discussion in Section 5.2)
and there is little information on the spatial
distribution of the water emission.  

There have been a few attempts to map the extended emission of water.
Gensheimer, Mauersberger, \& Wilson (1996) 
mapped emission of the quasi-thermal $3_{13} \rightarrow 2_{20}$ transition
of \HtwoeiO\ in both the Orion Hot Core and Sgr B2, and found that the emission
originated in a compact region ($< 10''$).  Cernicharo et al. (1994) and
Gonzalez-Alfonso et al. (1995) find evidence
for widespread water emission in Orion and W49N using the
$3_{13} \rightarrow 2_{20}$ masing transition of \HtwoO .   In Orion they  
argue that the water abundance is quite high, $x(\rm{H_2O}) > 10^{-5}$, over an extended 
$50'' \times 50''$ region centered on BN-KL and the Orion Nebular Cluster.   
From these results, the cloud-averaged water abundance for the central regions 
of the Orion core near the Orion Nebular Cluster is $^{>}_{\sim} 10^{-5}$ which,
using Figures 11 and 12 (\nhtwo\ $= 10^6$ \cc ; \cite{BSG96}), is consistent with 
$\tau_s \sim 10^6$ years, as suggested in Section 3.
However, the determination of abundances from masing transitions is a
difficult task and these results must be viewed as suggestive until they
are confirmed by mapping data obtained in other \HtwoO\ lines. 

For molecular oxygen, which also suffers from strong absorption due to atmospheric 
\Otwo , the situation is even more uncertain.  There exists only one tentative detection
of $^{16}$O$^{18}$O towards L134N by \cite{PLC93}, which implies $x(\rm{O_2}) 
\sim 4 - 8 \times 10^{-5}$.  However, a search for $^{16}$O$^{18}$O emission in 
different positions in L134N and other galactic sources by \cite{MPLC97} did not
confirm this detection and
provides only upper limits of \Otwo /CO $< 0.1$, which, assuming $x$(CO) $= 2.7 \times
10^{-4}$ (Lacy et al 1994), gives $x(\rm{O_2}) < 3 \times 10^{-5}$.  Searches, with 
similar negative results, have
also been performed for molecular oxygen in extra-galactic sources with favorable 
redshifts (Goldsmith \& Young 1989; Combes et al. 1991; Liszt 1992) 
finding $x(\rm{O_2}) <  10^{-6}$.
A recent study by Combes, Wiklind, \& Nakai (1997) towards
a z = 0.685 object provides the lowest limit to date of \Otwo /CO $< 2 \times 10^{-3}$.
For our most realistic case, the Monte Carlo gas-grain model, we find the time-averaged
\Otwo\ abundance is $\sim 2 \times 10^{-5}$.  Thus, the combined effects of
shocks and gas-grain chemistry could lower \Otwo\ abundances below the observed
limits in Galactic sources, but another explanation is required to account
for extra-galactic observations.  The low time-averaged \Otwo\ abundance suggests
that these results could have some bearing on the high abundances of neutral carbon
relative to CO (C/CO $\sim$ 0.1)
that are observed towards a variety of star forming regions (c.f. Plume 1995, Schilke
et al 1996).  However, because the number of dissociative shocks which destroy CO is 
not high enough, the time-averaged carbon abundance in the models is well
below the observed value. 

The best opportunity to test these results will be offered by two spaceborne observatories,
{\em The Submillimeter Wave Astronomy Satellite (SWAS)} (\cite{Melnick_etal95}) and 
{\em ODIN} (\cite{Hjalmarson95}), 
both of which should launch within the next year.   {\em SWAS} and {\em ODIN}
are capable of observing and mapping the fundamental transition
$1_{10} \rightarrow 1_{01}$ of \HtwoO\ at 557 GHz and the
$3,3 \rightarrow 1,2$ transition of \Otwo\ at 487 GHz.   
These transitions have low upper state energies ($\sim 26$ K)
and should be readily excited in hot gas near star-forming sites and,
more importantly, in the colder more extended material such as the ridge
of dense gas in Orion (c.f. \cite{Ungerechts_etal97}).  

\begin{figure*}
\figurenum{16}
\plotfiddle{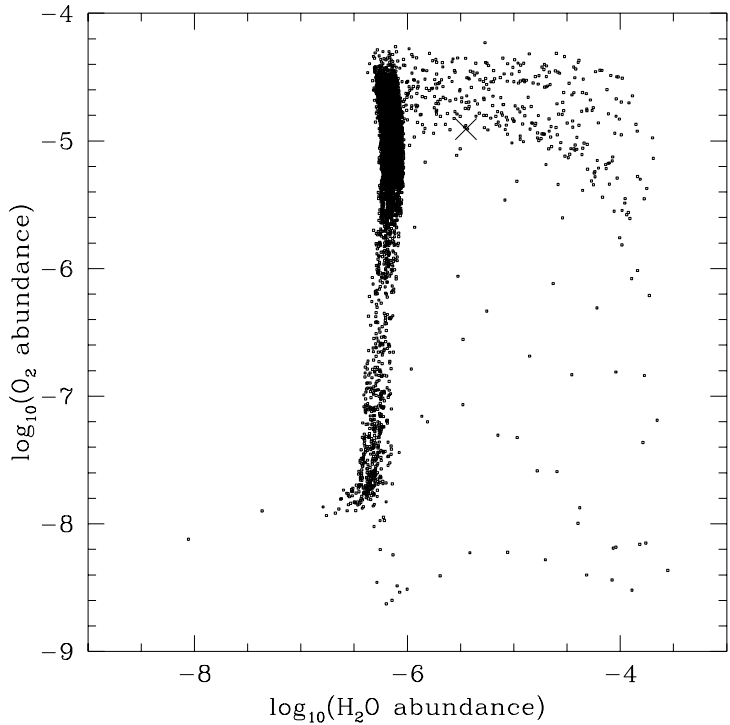}{2.5in}{0}{50}{50}{-160}{-75}
\caption{
Points defining the plane of O$_2$ and H$_2$O abundances for the 
gas-grain Monte-Carlo model with
$\tau_s = 10^{6}$ years and $n_{H_2} = 10^5$ cm$^{-3}$.  The spread of plotted 
points has been artificially increased by smearing them by 0.2 dex so as to give
a better sense of the density of points in different regions of the plot.  The cross
represents the mean (time-averaged) abundances. 
}
\label{shockspec}
\end{figure*}
Besides the computation of averaged abundances from mapping observations,
combined observations of \HtwoO\ and \Otwo\ should be a powerful tool in
examining the shock history of gas. 
This is demonstrated in Figure 16 where the water abundance is plotted
versus the molecular oxygen abundance for the Monte-Carlo model,
including grain depletion and desorption (Section 4.4.2).  In this plot,
the majority of points trace an area of roughly constant water
abundance of $x(\rm{H_2O}) \sim 10^{-6}$,
with the \Otwo\ abundance ranging between $x(\rm{O_2}) = 3 - 30 \times 10^{-6}$.
This area has, by far, the largest number of points and  
defines the ``main-sequence'' of quiescent chemistry.  From this main sequence
a shock will trace a line extending almost horizontally to the right, 
with constant \Otwo\ abundance.  Very strong shocks continue at the end of the horizontal
line and extend down  
almost vertically, lowering the \Otwo\ abundance for constant \HtwoO\ abundance.
This dependence of horizontal and vertical lines occurs because the \HtwoO\ is created
more efficiently than \Otwo\ destruction (see Figures 3 and 4).
This plot is a different way of examining the models, as opposed to average abundances,
because it presents the evolution in a continuous fashion; the various
possible solutions define a plane that traces evolutionary tracks including quiescent
chemistry through a broad range of shock strengths.

\subsection{Post-Shock Chemistry and Water Ice Mantles}

For comparison with previous modeling efforts,
our three-stage model is similar to the episodic models presented in Charnley
et al. (1988a,b).  However, our model is simpler in that we do not
model the formation process of a dense clump nor do we account for any mixing between
shocked and non-shocked layers.  Like their models, we find that the chemistry
converges to well defined abundance values even when numerous cycles are repeated.
Charnley et al. (1988a) also find that 
the water abundance varies greatly between shock and quiescent cycles, but 
they do not examine in detail the chemistry in the post-shock layer.

An interesting result in the gas-grain models is that the abundance of water on
grain surfaces in post-shock gas is quite large $x(\rm{H_2O})_{gr} \sim 10^{-4}$. 
This abundance is quite close to that inferred for water ice along lines of sight
towards background stars in Taurus, where $x(\rm{H_2O})_{gr} \sim 8 \times 10^{-5}$ 
(\cite{Whittet93}).   Thus, these results offer an alternative explanation for the large
abundance of water in grain mantles -- one that requires no grain surface chemistry!  
We stress that this mechanism may not be responsible for all \HtwoO\ 
observed on grains.  Grain surface chemistry formation of \HtwoO\ must be
considered -- especially during cloud formation stages when the abundance of
atomic hydrogen and oxygen is higher.  
There are also other potential desorption mechanisms that could be active
and are not included in these models, such as grain mantle explosions induced
by UV radiation (\cite{SG91}) or water desorption via the infrared radiation field (\cite{WHW92}). 
However, since most molecular clouds are very active star-forming sites,
we suggest that caution must be applied when interpreting
observations of ice mantles in molecular clouds as the sole result of grain surface
chemistry.     
The interpretation of high gas-phase water abundances towards
hot stars as the result
of either high-temperature chemistry or grain mantle evaporation of \HtwoO\ is further
obscured because water on
grains may have been produced in an earlier shock episode.  It is possible that
the HDO/H$_2$O ratio could discriminate between water mantles
created in shocks or in grain mantles and  we are in the process of 
investigating this question (Bergin, Neufeld, \& Melnick 1997). 

The large abundance of water on grains, with CO remaining in the gas phase,
will also alter the ratio
of total carbon-to-oxygen (C/O) in the gas phase.  Comparison of chemical theory
with observations of molecular abundances in GMC cores associated with
massive star formation has found that 
the chemical abundances of many species are 
best reproduced with C/O ratios greater than the
solar value ($>$ 0.4; \cite{BSMP87}; \cite{BGSL97}).  
For chemical modeling this is the result of depletion of the initial abundance of
oxygen relative to carbon, which reduces
the abundances of small oxygen-bearing species that are major carbon destroyers
in the gas phase.    In the three-stage gas-grain model shown in Figure 7, the C/O
ratio in the post-shock gas at $t \sim 10^{5}$ years is C/O $\sim 0.7$.  It is
difficult to gauge the effect of this on other species, because of the lack of 
certainty with regard to high-temperature reactions.  
It is therefore possible that the C/O ratios inferred in 
Blake et al. (1987) and Bergin et al. (1997) are indicative of
the core formation process, which could 
involve gas temperatures rising high enough to convert
atomic oxygen to water.  
In this case, these results suggest a simple mechanism to place
large amounts of oxygen on grain surfaces while still keeping most of the
carbon in the gas-phase.  

Other molecular species, such as SO, SiO (\cite{MPBF92}), 
and \CHthreeOH\ (\cite{BLWC95}) 
have been observed with enhanced abundances in energetic outflows (see also \cite{Bachiller96}; \cite{Bachiller_Perez97}).   
These species are included in the three-stage chemical model presented in Section 4.3, but
we do not present the results in detail here 
because of uncertainties in the high-temperature
reaction rates.  However, these results do have some bearing on the chemistry of
these species in the post-shock layer.   The formation pathway of 
\CHthreeOH\ in the gas phase is linked to a reaction of CH$_3^+$ with \HtwoO\
(\cite{MHC91}).  Thus,  it is possible that the enhanced 
abundance of water through high-temperature reactions could lead to larger
\CHthreeOH\ abundances through this reaction.  In the three-stage
model, the \CHthreeOH\ abundance in the post-shock layer is $x(\rm{CH_3OH}) = 1 - 5 \times
10^{-8}$ when the shock temperature ranges from 1000 to 2000 K. 
These abundances are at least an order of magnitude below values inferred in
molecular outflows, which can be as high as $x(\rm{CH_3OH}) \sim 10^{-6}$ (\cite{BLWC95}). 
The mechanism of methanol enhancements
may therefore be the result of grain mantle evaporation
or an unidentified high- or low-temperature pathway.  The abundance of \CHthreeOH\ accreted
onto grain surfaces provides another constraint.  In our model the abundance of
\CHthreeOH\ ice in the grain mantle is only 0.1 percent of the water-ice abundance.
This ratio is below that observed towards NGC 7538 or W33A where $x(\rm{CH_3OH})$/$x(\rm{H_2O}) 
\sim 10-40$ percent (\cite{ASTH92}).  Our models do not set constraints on
the chemistry of SO and SiO because the large abundances of these species in outflows may
be the result of sputtering of grain refractory and/or mantle material 
(c.f. Schilke et al. 1997; Caselli et al. 1997).  

When the 
abundances of these species are enhanced, through any mechanism, the high abundances 
will persist until the timescale for the individual molecule to deplete 
onto grain surfaces.
If the dust temperature is high enough to
keep a given species in the gas-phase, or for pure gas-phase
chemistry, the lifetime will
be $\tau_{ra} = 4 -7 \times 10^{5}$ years.
In these models all SiO depletes onto grain surfaces
at $\tau_{dep}
\sim 10^4$ years (for \nhtwo\ = 10$^{5}$ \cc\ and assuming a sticking
coefficient of unity).  For \CHthreeOH\ abundance enhancements we find that 
the depletion timescale for gas-grain chemistry
is equal to that of \HtwoO\ ($\tau_{dep}(H_2O) \sim 10^5$ years). 
The disparity in post-shock lifetimes between \CHthreeOH , \HtwoO , and SiO suggests
that differences might exist between younger outflows, perhaps those associated with
so-called Class~0 sources, and the more evolved sources (e.g. Class I). 
A survey of such sources in these tracers might prove to be useful in 
probing the links between evolutionary effects observed in outflows and the
relationship to the driving source. 
Another possibility is that differences could also be found within different
components inside a single outflow  (see \cite{Bachiller_Perez97}). 

The abundance of some molecular species, notably \HCOp\ and \NtwoHp , are adversely affected 
by the high water abundance.  In Figure 7, for the three-stage model, the abundance of
\HCOp\ is decreased through reaction R6 when the water abundance is raised.  
Thus, the abundances of these
two important molecular species should be anticorrelated. 
Because \NtwoHp\ also reacts with \HtwoO , a similar 
anticorrelated behavior would be found  between the abundances of \NtwoHp\ and \HtwoO\ and
may account for the low \HCOp\ and \NtwoHp\ abundances in hot regions such as the Orion
Hot Core (c.f. Blake et al. 1987) and in the L1157 outflow (\cite{Bachiller_Perez97}).

\section{Summary}

We have investigated the use of simple oxygen-bearing molecules, 
\HtwoO\ and \Otwo , as tracers of the shock history of molecular gas.   
To this end, we have constructed a coupled chemical and dynamical model of
dense molecular gas that is subject to heating by non-dissociative
C-type shocks, and subsequently cools. 
To constrain the effects of MHD shocks on molecular gas we use the 
shock models of Kaufman \& Neufeld (1996a,b).   We have constructed two models
and concentrate, in particular, on the chemistry in a layer of post-shock
cooled gas. 
In the first model we examine the chemical evolution in three-stages: (1) 
as molecules form and increase in abundance within quiescent, pre-shock
gas; (2) the rapid chemical changes precipitated by the passage of a
C-shock; and, (3) the post-shock chemical evolution that occurs as the
gas cools.  The results of the three-stage model are combined with
observational estimates of shock rates within molecular gas in regions
of molecular outflows.   
A second, more general, model uses a Monte-Carlo method to 
examine the chemistry in a parcel of gas that is subject to random
shock heating over a cloud lifetime.   
Because water is also an abundant molecule on the surfaces of interstellar
grains, we have separately examined the pure gas-phase chemistry and
the gas-phase chemistry including grain surface molecular depletion
and desorption.

The principal results from this study are:

1) For $v_s > 10$ km s$^{-1}$, high temperature neutral-neutral
chemical reactions are efficient at converting all of the available
oxygen into \HtwoO\ within the post-shock region before the gas has
an opportunity to cool.  For lower shock velocities, the water abundance
will not change significantly from
its quiescent values due to the rapid post-shock cooling to temperatures less than those
required to initiate these neutral-neutral reactions.
This result is in agreement with previous investigations.
Similarly, for $v_s > 26$ km s$^{-1}$, all \Otwo\ will be destroyed inside the
shock within the cooling time.

2)  For pure gas-phase models, the enhanced \HtwoO\ abundance behind a shock
will decrease and the low-temperature 
chemistry will be re-asserted at $\tau_{ra} \sim 4 - 7 \times 10^5$ years.  
As a result of
cosmic-ray driven chemistry, $\tau_{ra}$ is nearly independent of the
gas density.
For models which include gas-grain interactions, 
a density dependence is found as the large abundance of water in the gas-phase
will decrease though accretion onto grain surfaces at the depletion timescale.  
The depletion timescale has an inverse dependence on the density 
and is less than $\tau_{ra}$ for \nhtwo\ $> 10^{5}$ \cc .

3) We find that the time-averaged abundance of \HtwoO\ is a sensitive function
of the shock frequency.  This abundance is approximately the weighted
average of the amount of time gas spends in the pre-shock, shocked, and
post-shock stages respectively, and determines the average abundance of water
mapped over a large region. 
Thus these models predict that the abundance of \HtwoO , and to a lesser
extent \Otwo, can be used to constrain the physical history of molecular gas.
The observations required to test these predictions can be performed
from {\em The Submillimeter Wave Astronomy Satellite} and {\em ODIN}. 
 
4) The abundance of water-ice on grain surfaces can be quite large
in the post-shock gas layer.  The abundance is often $x(\rm{H_2O})_{gr} > 10^{-4}$ 
and is comparable to that observed in molecular clouds.  This offers an
alternative method to create water-ice mantles without resorting to
grain surface chemistry -- passage of a C-type shock, followed by rapid
hydrogenation of gas-phase oxygen, and subsequent depletion of water onto cold
dust grains.   Despite the large amount
of water trapped on grains, the gas-phase chemistry is maintained
because CO and \Otwo\ do not deplete.  This differential depletion
of \HtwoO\ compared to CO and \Otwo\ can also effectively raise
the C/O ratio in the gas phase to values that are inferred from previous
observational and theoretical comparisons of chemical abundances in GMC cores.   

5) The abundance of a few molecular species are adversely affected by the high water abundances in
shocked regions.  In particular the abundances of \NtwoHp\ and \HCOp\ are predicted
to have abundances that are anticorrelated with changes in the water abundances.  

We thank the referee R. Bachiller for a thorough review and some helpful comments.
We also thank C. Lada and R. Snell for useful discussion on shock timescales in 
molecular clouds, and A. Dalgarno for discussions on chemical reactions. 
We are also indebted to M. Kaufman for kindly providing output from his
shock models.
E.A.B. and G.J.M. acknowledge the support of NASA grant NAGW-3147 from the 
Long-Term Space Astrophysics Research Program; and D.A.N. acknowledges the support
of the Smithsonian subcontract SV-62005 from the SWAS program.

\end{document}